\documentclass[pre,showpacs,11pt,amsfonts,amsmath,amssymb]{revtex4}

\usepackage{amssymb}
\usepackage{graphicx}
\usepackage{dcolumn}
\usepackage{bm}
\usepackage{appendix}

\newcommand{\EE}[1]{\mathbb{E}\left[#1\right]}

\newcommand{\cI}{{\mathbf{1}}}
\newcommand{\bI}{{\overline{\mathbf{1}}}}
\newcommand{\cC}{{\cal C}}
\newcommand{\cA}{{\cal A}}
\newcommand{\iI}{{\cal I}}
\newcommand{\cS}{{\cal S}}
\newcommand{\cT}{{\cal T}}
\bibstyle{revtex}

\begin{document}

\preprint{APS/123-QED}

\title{Self-similar continuous cascades supported by random Cantor sets. Application to rainfall data}

\author{Jean-Fran\c{c}ois Muzy}
\email{muzy@univ-corse.fr}
\affiliation{SPE UMR 6134, CNRS, Universit\'e de Corse, 20250 Corte, France}
\affiliation{CMAP UMR 7641, CNRS, Ecole Polytechnique, 91128 Palaiseau, France}
\author{Rachel Ba\"{\i}le}
\email{baile@univ-corse.fr}
\affiliation{SPE UMR 6134, CNRS, Universit\'e de Corse, 20250 Corte, France}

\date{\today}

\begin{abstract}
We introduce a variant of continuous random cascade models that extends former 
constructions introduced by Barral-Mandelbrot \cite{BaMan02} and Bacry-Muzy \cite{MuBa02,BaMu03}
in the sense that they can be supported by sets of arbitrary fractal dimension. 
The so introduced sets are exactly self-similar stationary versions of random Cantor sets formerly
introduced by Mandelbrot as ``random cutouts'' \cite{Man72}.
We discuss the main mathematical properties of our construction 
and compute its scaling properties. We then illustrate our purpose on several numerical examples and we consider
a possible application to rainfall data. We notably show that our model allows us to reproduce remarkably the
distribution of dry period durations.
\end{abstract}


\maketitle
\section{Introduction}
Random multiplicative cascades were introduced by the Russian
school \cite{novikov} in order to describe the energy transfer from large
to small scales in fully developed turbulence \cite{Fri95}.
Theses processes, further studied by B. Mandelbrot \cite{Man74a,Man74b} 
(and usually referred to as ``Mandelbrot Cascades'') are built using an iterative rule that spreads some measure (which can represent mass, energy,...) 
from large to smaller scales: 
the measure of the two half sub-intervals of some given interval is obtained 
by multiplication of its mass by two independent random factors which 
law remains the same at all scales.
Mandelbrot cascades were also generalized to measures
supported by sets of non-integer dimension (see below).
The main problem with such ``grid bound'' random cascades is that they are 
always constructed on a bounded interval (the starting largest scale),
they are not stationary processes and they necessarily involve some 
preferred scale ratio. 
In that respect, they can hardly be used to reproduce faithfully natural 
phenomena like the energy dissipation in fully  
developed turbulence or the volatility in financial markets.

The first continuous extension of Mandelbrot
cascades was proposed a decade ago by Barral and Mandelbrot \cite{BaMan02}.
These authors replaced the regular tree-like construction underlying
previous iterative scheme by a ``random tree'' associated with the
draw of a Poisson process in the ``time-scale'' half-plane.
Later, Bacry and Muzy \cite{MuBa02,BaMu03} extended this construction by replacing the compound Poisson law by any infinitely divisible law.
The main advantage of continuous random cascades is that they
provide a large class of models that are stationary and possess continuous
scale invariance properties. They can be considered as the paradigm
of random multifractal processes and they have found applications
in many domains like turbulence or empirical finance  
(see e.g ref.\cite{BacKozMuz06} for a recent review). 

The goal of this paper is to construct models of continuous cascades that
are not necessarily supported by the whole real line \footnote{For the sake of simplicity we will restrict our considerations to one-dimensional models. Extension to 
higher dimensional spaces is not straightforward and will be considered in a future work.}
but by sets of zero (Lebesgue) measure. For that purpose, 
we will introduce a family of stationary random 
self-similar Cantor sets, i.e., closed sets that contain
no interval \cite{Falc03}. These sets turn out to be a particular version
of the construction proposed by Mandelbrot in early seventies
that consists in cutting the real line ``at random'' with some given 
distribution of ``cutouts'' \cite{Man72}. We show this self-similar version can be the support of a continuous log-infinity divisible random cascade. We study the main properties of the proposed model and consider, 
as a first application, the statistical modeling of precipitations.

The paper is organized as follows: in section \ref{sec:casc} we recall how discrete and continuous 
cascades are defined and how a time-scale approach can be conveniently used to go from the former to the 
later. Along the same line, we study in section \ref{sec:cantors}, how classical deterministic and discrete
constructions of Cantor sets could be extended to a continuous scale framework. This leads us to rediscover
the Mandelbrot random cutouts and to introduce a version that is exactly self-similar. We study the fractal dimension
of these random sets and the statistics of their void size. The general model, namely a self-similar multifractal random measure
lying on a random Cantor set is introduced in section \ref{sec:meas}. We show that the model is well-defined (i.e. the construction
converges to a non trivial limit) and is indeed self-similar in a stochastic sense with (multi-)scaling properties that 
extend those of continuous cascades to non-integer dimensions. We provide various examples in section \ref{sec:examples}
and consider an application to account for the intermittency of precipitation rates. In particular, we show that our approach
allows one to reproduce very well the statistics of the dry period lengths. 
Concluding remarks are provided in section \ref{sec:conclusion} 
while the more technical parts of this paper are reported in appendices. 

\section{From discrete to continuous cascade models}
\label{sec:casc}

A convenient way to describe the construction of a Mandelbrot cascade over the interval $[0,T]$ 
is the following:
Let $\cI_{\cA}(t)$ denote the indicator function of the set ${\cal A}$ and let us the function: 
\begin{equation}
	\label{defP}
	\bI_\cA(W,t) = 1+(W-1)\cI_{\cA}(t)
\end{equation} 
One thus has $\bI_\cA(W,t)=W$ if $t \in \cA$ and $\bI_\cA(W,t)=1$ otherwise.
The Mandelbrot random cascade is the weak limit, when $n \to \infty$, of 
the sequence of densities:
\begin{equation}
\label{GB-casc}
  \sigma_n^M(t) = \prod_{i=1}^{n} \prod_{j=1}^{2^i} \bI_{\iI_{ij}}(W_{ij},t)
\end{equation}
where $0 \leq t \leq T$, $\iI_{ij}$ denotes the dyadic intervals $[jT2^{-i},(j+1)T2^{-i})$ and
$W_{ij}$ are independent copies of a positive random variable such that $\EE{W}=1$.
Kahane and Peyri\`ere \cite{KahPey76} proved that such a sequence has almost surely
a non-trivial limit provided $\EE{W \ln W} < 0$.
\begin{center}
	\begin{figure}[h]
		\includegraphics[width=0.5 \textwidth]{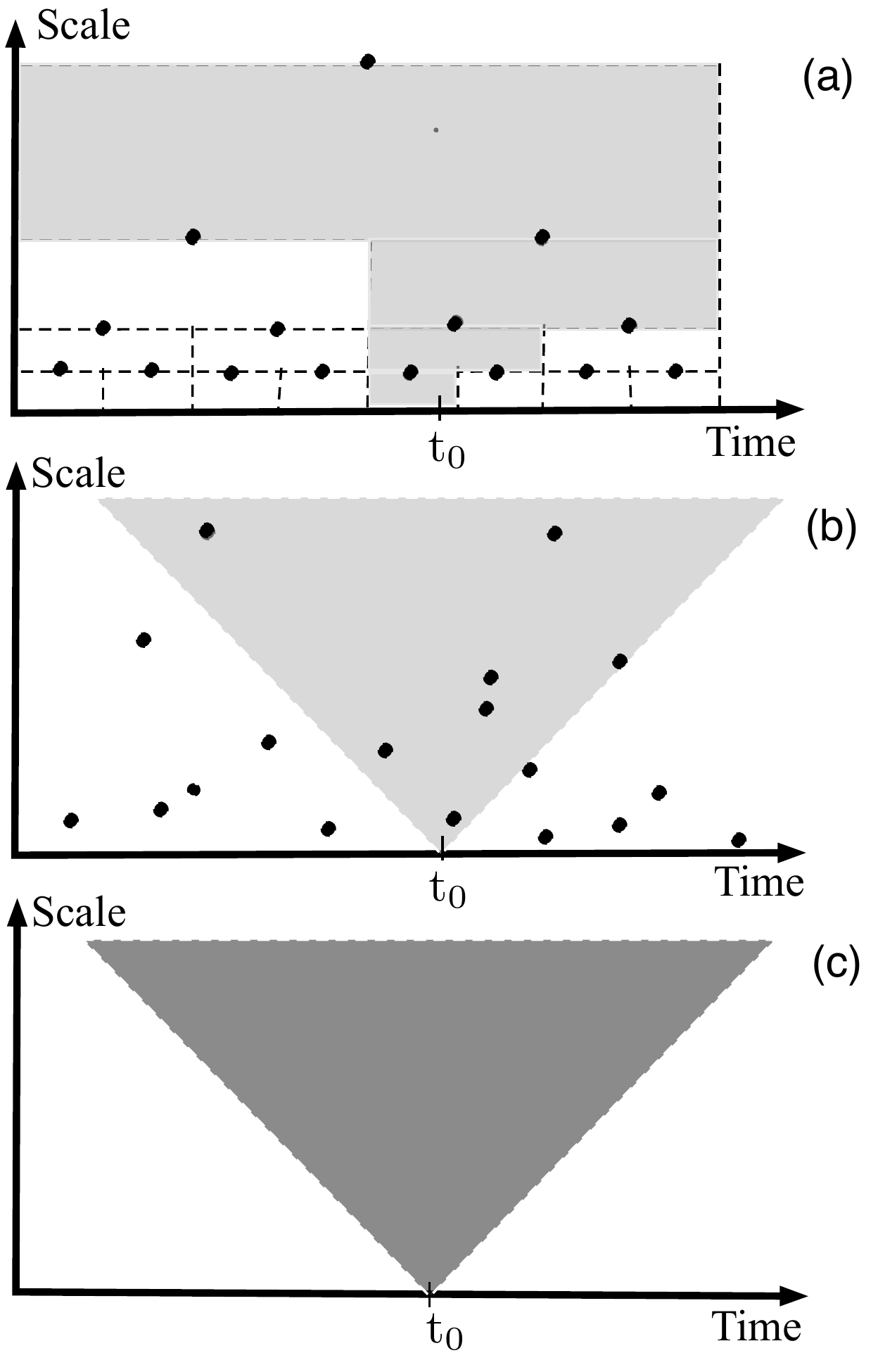}
		\caption{From grid bound to continuous random cascades: (a) Sketch of Mandelbrot cascade construction: the
			measure at some given point $t_0$ is obtained as the product of random factors $W_i$ associated 
			with the 
			dyadic intervals of size $T 2^{-i}$ which contain $t_0$. (b) Barral-Mandelbrot Cascade: the measure at $t_0$ is now obtained as the product of random weights $W_i$ associated with the points $(t_i,s_i)$ 
			of a Poisson process in the time-scale plane such that $t_0 \in \iI_{t_i,s_i}$. This corresponds to the events 
			$(t_i,s_i)$ that are in the cone $\cC(t_0)$. (c) Bacry-Muzy log-infinitely divisible extension: The measure at $t_0$ is simply written as $e^{\omega_\ell(t_0)}$ where $\omega_\ell(t)$ is the integral over the cone
			$\cC_\ell(t_0)$ of a random infinitely divisible noise.}
		\label{fig:casc}
	\end{figure}
\end{center}

Barral and Mandelbrot proposed a stationary, ``grid free'' version of cascade models by replacing the dyadic grid (represented by the dyadic intervals in Eq. \eqref{GB-casc}) 
with a Poisson process in the time-scale plane (see Fig. \ref{fig:casc}).
More precisely, let $(s_i,t_i)$ be the events of a Poisson process of intensity
$c s^{-2} dt ds$ \footnote{Indeed, as advocated in \cite{MuBa02,BaMu03}, this measure
is invariant under the dilation-translation group acting on $(s,t)$ (it corresponds to the left Haar measure).}.
With each couple $(t_i,s_i)$ one can associate a time interval 
\begin{equation}
\label{timeinterval}
\iI_{t_i,s_i} = \left(t_i-\frac{\min(T,s_i)}{2},t_i+\frac{\min(T,s_i)}{2}\right)
\end{equation}
where $T > 0$ is a largest interval size (the so-called ``integral scale'' in turbulence cascade models).
Let $W_i$ be i.i.d. copies of a 
positive random variable such that $\EE{W} = 1$. The Barral-Mandelbrot measure of integral scale $T$ 
is then defined as the (weak) limit, when $\ell \to 0$, of the density \footnote{In the original construction of Barral and Mandelbrot
the large scale $T$ is not introduced in that way but the definition below corresponds to an exactly self-similar version.}:
\begin{equation}
\label{BMC}
   \sigma^{BM}_\ell (t) = \prod_{(s_i,t_i),s_i \geq \ell} \bI_{\iI_{t_i,s_i}}(W_i,t)
\end{equation}

Barral-Mandelbrot construction was extended by Bacry and Muzy \cite{MuBa02,BaMu03}
that proposed to replace the (compound) Poisson measure in the time-scale plane
with any infinitely divisible random measure $dm(s,t)$.
The extension relies on the the following remark: if one denotes
by $\cC_\ell(t_0)$ the domain in the time-scale plane such that, for
each $(t,s) \in \cC_\ell(t_0)$, $t_0 \in \iI_{t,s}$, then
Eq. \eqref{BMC} can be rewritten as:
\begin{equation}
\sigma^{BM}_\ell (t) = e^{\sum_{(t_i,s_i) \in \cC_\ell(t)} \ln(W_i)} = e^{\int_{\cC_\ell(t)} dp(t,s)}  \;, 
\end{equation}
where $dp(t,s)$ is the Poisson measure of intensity $c s^{-2} dt ds$ compound
with the random variable $\ln W$. Bacry and Muzy suggest to replace the 
compound Poisson law with a general infinitely divisible law $dm(s,t)$
and consider the weak limit of the density (see Appendix \ref{cone_app}):
\begin{equation}
  \sigma_\ell(t) =  e^{\omega_\ell(t)} = e^{\int_{\cC_\ell(t)} dm(s,u)} 
\end{equation}
This is illustrated in Fig. \ref{fig:casc}.
In the following we will denote by $\mu(t)$ the so-defined 
limit measure, i.e.
\begin{equation}
  \mu(t) = \lim_{\ell \to 0} \int_0^t \sigma_\ell(u) du
\end{equation}

\section{Stationary random Cantor sets: the Mandelbrot random cutouts}
\label{sec:cantors}

\subsection{A stationary version of the ``middle-third'' Cantor set}
The previous framework allows one to construct stationary random measures that are supported by the whole real line.
As discussed in the introductory section, it could be interesting to extend these constructions to situations
where the measure is supported by a set of non-integer dimension.
The paradigm of such set is the so-called Cantor set, and notably
the ternary Cantor set (also called ``triadic'' or ``middle third'' Cantor set) that can be constructed
as follows: 
one starts with some interval that can be chosen
to be $\iI_0 = [0,1]$ with no loss of generality.
At the first step, one deletes the middle third open interval $(1/3,2/3)$
to obtain $\iI_1 = \iI_{11} \cup \iI_{12} = [0,1/3] \cup [2/3,1]$.
The operation is then repeated on $\iI_{11}$ and $\iI_{12}$  which middle third 
interval is deleted and so on {\em ad infinitum}. One can show that the limit set $\cT$, i.e., the set of points
of $[0,1]$ that are not removed by the previous iterative construction,
exists and is non trivial. It is a so-called ``Cantor set'' 
in the sense that it is a closed set of uncountable cardinality that
contains no interval. 
It is well known that $\cT$ is a self-similar set of zero length (Lebesgue measure) and of fractal dimension $d_C = \frac{\ln 2}{\ln 3}$.

Let us notice that various random 
extension of the previous construction have been considered in the literature by, for example, picking randomly the scale of the subdivision at each construction step ($1/3$) or the position of the deleted interval. 
One of the most famous random Cantor model is the so-called $\beta$ 
model introduced by Frisch, Sulem and Nelkin \cite{FSN}.
At each step of the division (by e.g a factor 2), one keeps each subinterval
with a probability $\beta$ and one removes it with a 
probability $1-\beta$. If $d_C$ stands for the 
fractal dimension of the so-obtained limit set,
at step $n$ (and thus at scale $2^{-n}$), it remains in average
$2^{n d_C} = 2^n \beta^n $ subintervals. The dimension $d_C$ 
is therefore $d_C = 1+\log_2(\beta)$. However, 
for the same reasons that grid bound cascades 
are not convenient for most physical applications, these approaches are not satisfactory: they do involve a preferred scale ratio and are not invariant by time translation. It is then natural to address the question
of their extension to a ``continuous'' and stationary version. 

The first attempt to solve this problem can be found in ref. \cite{Schmitt98} where
the author considered a ``continuous'' $\beta$ model by letting the scale ratio 
at each construction step going to $1$ and considering in the same time 
$\beta \to 1$ in such a way that the dimension $d_C$ remains constant. 
If this approach is formally appealing, it remains hard
to handle mathematically since a scale ratio that approaches $1$
corresponds to strongly overlapping subintervals. 
In a more recent paper \cite{Schmitt14}, Schmitt proposed an extension of the $\beta$ model in the framework of log-infinitely divisible distributions as described in the previous section. Although this paper proposes a very interesting 
path towards continuous Cantor sets and can be directly related to our construction, the author did not considered the existence of the small scale limit of his model and neglected the necessary time correlations
in the random factors he introduced (see Section \ref{sec:mf}).
 
\begin{center}
	\begin{figure}[h]
		\includegraphics[width=0.5 \textwidth]{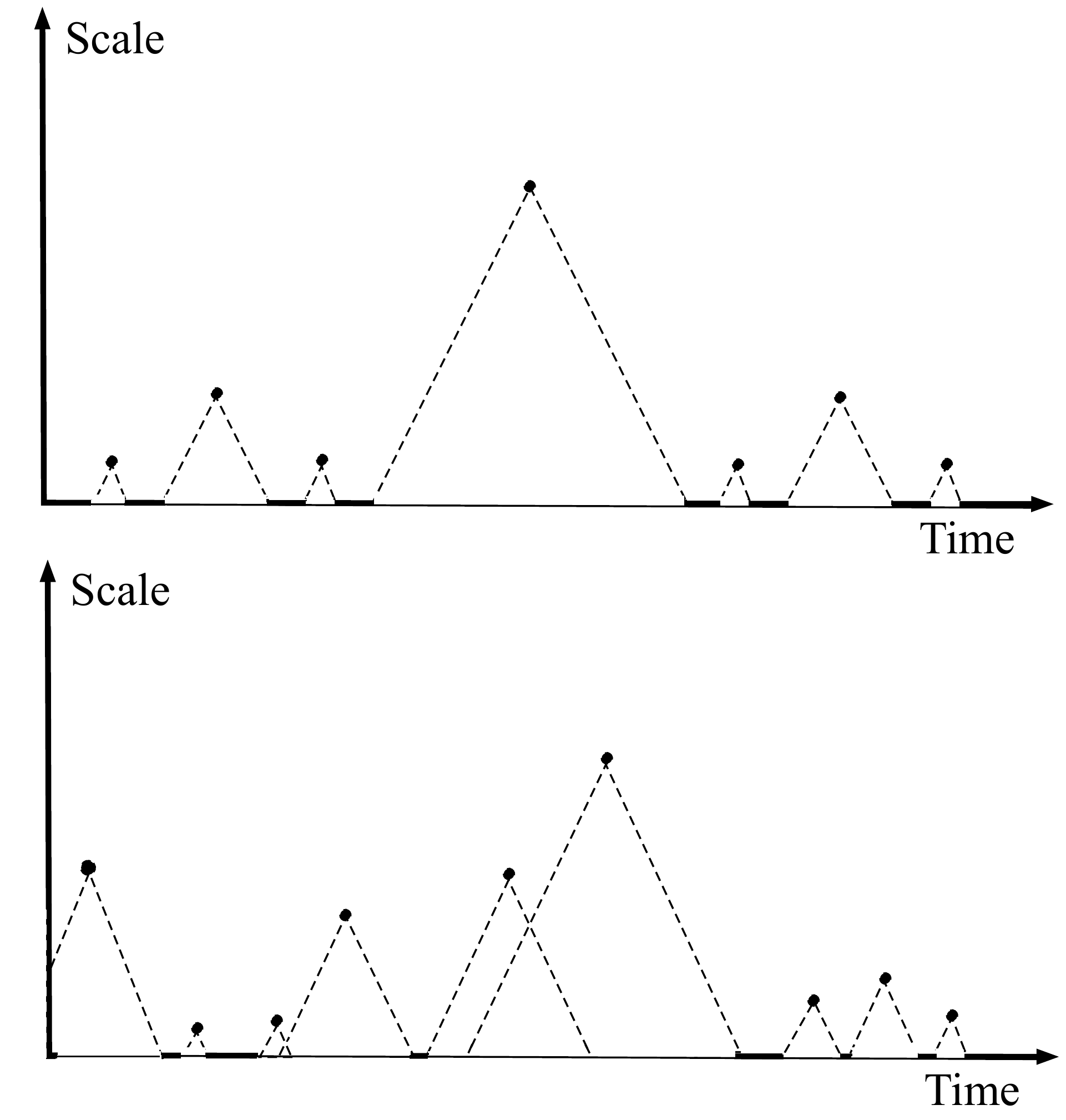}
		\caption{Illustration of classical vs continuous Cantor construction:
			On the top panel the construction of the deterministic triadic Cantor
			set is depicted: one chooses precisely the points in the time-scale half-plane $(t,s)$ along the regular grid
			at scales $s_i = 3^{-i}$. In the bottom panel, the points are drawn from a Poisson law with the same mean density 
			as in the deterministic example. The resulting random set has the same dimension as in the deterministic case.}
		\label{fig:cantor}
	\end{figure}
\end{center}

In order to go from discrete (i.e. grid bound) Cantor sets to their continuous analog,  let us first remark that, within the representation introduced of Sec. \ref{sec:casc} 
for continuous cascades \cite{BaMan02,BaMu03}, the construction of the
ternary Cantor set $\cT$ can be redefined as follows: if one considers for some given point $(s,t)$ in the time-scale plane, the interval
$\iI_{t,s} = (t-s/2,t+s/2)$ centered at $t$ and of size $s$, then $\cT = \lim_{n \to \infty} \cT_n$ with
\begin{equation}
 \cT_n = [0,1] \setminus \bigcup_{m=1}^n \; \; \bigcup_{t_k \in E_m} \iI_{t_{k},s_m}
\end{equation}
where $s_m=3^{-m}$ and $E_m$ is the set of $2^{m-1}$ points defined by the recurrence $E_1 = \{1/2\}$ 
and $E_m = \frac{1}{3}E_{m-1} \cup \frac{2}{3}+\frac{1}{3} E_{m-1}$.
The set $\cT_3$ is depicted in the top panel of Fig. \ref{fig:cantor} where we have represented
the locations of the points $(s_i,t_{k})$ in the half-plane $(s,t)$ associated with its 3 components.

A natural way to ''randomize'' this definition in a fully stationary way 
is then to follow the same approach as 
in the Barral-Mandelbrot generalization of grid bound cascades:
One simply replaces the fixed locations $(s_m,t_{k})$ in the $(t,s)$ half-plane with random locations drawn using a Poisson point process of intensity chosen in order to preserve the density of the triadic grid.
Let $dp(t,s) = K s^{-2} dt ds$ be the intensity of a Poisson process homogeneous with respect to the natural measure of the time-scale half-plane.
One seeks for $K$ such that
the density of points in the Poisson case corresponds to the one of the ternary Cantor set. Since
at each generation, one deletes one third of the remaining intervals,
if $P_m$ is the probability to have no Poisson point within the domain $s \in [3^{-m},3^{-(m+1)}]$
and $t \in [t_0-s/2,t_0+s/2]$, one must have $P_m = 2/3$.
From Poisson distribution properties, we have $P_m = e^{-\lambda_m}$ where $\lambda_m$ is the intensity
of the number of Poisson points in the domain $s \in [3^{-m},3^{-(m+1)}]$, $t \in [t_0-s/2,t_0+s/2]$, i.e.,
 \begin{equation}
 \lambda_m = K \int_{3^{-(m+1)}}^{3^{-m}} \! \! \! \! s^{-2} ds \int_{t_0-s/2}^{t_0+s/2} \! \! \! \! dt = K \ln(3) \; .
\end{equation}
It thus results that
$-K \ln(3) = \ln(2)-\ln(3)$ ,i.e., $K = 1-d_C$ 
where $d_C = \frac{\ln 2}{\ln 3}$ is the fractal (capacity or Hausdorff) dimension of the Cantor set.

\subsection{Self-similar Mandelbrot random cutouts and their fractal dimension}
\label{cutouts}

The previous example suggests to introduce a new construction of a random Cantor $\cS$ supported
by the whole real line by removing random intervals drawn using a non-homogeneous Poisson process.
It turns out that this idea has already been proposed in a remarkable work by Mandelbrot 
in early seventies \cite{Man72}. Mandelbrot precisely raised the problem of a generalization 
of the middle-third Cantor set as we did in previous section 
and introduced the notion of ``random cutouts''. In doing so, 
he introduced the analog on the real line of a similar 
``cutout'' problem considered by Dvoretzky \cite{Dvor56} on the circle.
Mandelbrot studied the general case when the Poisson proceses in the $(t,s)$ plane is non-homogeneous 
along the scale direction and characterized by some intensity $\Lambda(s)$. He notably
studied some conditions on $\Lambda(s)$ under which one has $\cS = \varnothing$.
This problem was solved by Shepp \cite{Shepp72} that proved that,
with probability one, $\cS$ is non-empty if and only if
\begin{equation}
\label{shepp-condition}
\int_0^1 dx \; \; e^{\int_x^\infty (s-x) \Lambda(s) ds} < \infty
\end{equation}
while it is empty with probability one if this condition fails.

According to the considerations of the previous section, let us introduce
a non-homogeneous Poisson process in the plane $(t,s)$ which intensity  
$\Lambda(s) = 0$ for $s > T$ and reads, for $s \leq T$,
\begin{equation}
\label{intensity}
\Lambda(s) = (1-D) \left[ s^{-2}+ T^{-1} \delta(T-s) \right]
\end{equation}
where the parameter $D$ is such that $0 < D \leq 1$, $T>0$ is a large scale cut-off parameter (the ``integral scale'') and 
$\delta(t)$ stands for the Dirac function.
Let us note that bounding the range of $s$ to $s \leq T$ and 
adding this Dirac term is equivalent to consider that $\Lambda(s) = (1-D) s^{-2}$
for all $s > 0$ and to associate with each $s$ an intervall of size $\min(T,s)$, as in Eq. 
\eqref{timeinterval}. This trick allows one, as we will see below, to preserve 
exact self-similarity (see \cite{MuBa02,BaMu03}).
If $(t_i,s_i)$ are the events of such a Poisson process, 
a self-similar Mandelbrot cutout $\cS_D$  of parameter $D$ is defined as:
\begin{equation}
\cS_D  = \lim_{\ell \to 0} \; \; \mathbb{R} \setminus \bigcup_{s_i \geq \ell, t_i}  \iI_{t_{i},s_i}
\end{equation}

Using the above Shepp criterion, one can easily see that this set is non-empty with 
probability one. Indeed, from Eq. \eqref{intensity}, the integral \eqref{shepp-condition}
simply becomes:
\[
T^{1-D} \int_0^1 x^{D-1} \; \; dx
\]
which is always finite if $D>0$.

One can even get a stronger result and show that, with probability one, $D$ is the Hausdorff
dimension of $\cS_D$.
This directly results from the following theorem proven by Fitzsimmons et al. \cite{Fitz85}:
With probability one, $dim_H(\cS_D)$, the Hausdorff dimension $d_H$ of $\cS_D$ is 
\begin{equation}
 0 \vee \sup \left\{ \rho, \;  u^{1-\rho} e^{\int_u^1 dz \int_{z}^{\infty} \Lambda(s) ds } \to 0 \; \mbox{as} \; u \downarrow 0    \right\}
\end{equation}
Since the double integral in the exponential term is simply $(D-1) \ln(u)$, the Hausdorff dimension
is the greatest value of $\rho$ such that $\lim_{u \to 0} u^{D-\rho} = 0$, i.e., $d_H = dim_H(\cS_D) = D$.

\subsection{The void probability distribution and box-counting dimension of $\cS_D$ }
\label{sec:void}
The box-counting dimension $d_C$ (also referred to as ``covering" or ``capacity" dimension) of a set is defined 
from the behavior, when $\varepsilon \to 0$, of $N(\varepsilon)$, the number of boxes of size $\varepsilon$ necessary to cover this set:
\begin{equation}
 N(\varepsilon) \sim \varepsilon^{-d_C}
\end{equation}
This means that the probability that a given box 
$[t_0-\varepsilon/2,t_0+\varepsilon/2]$ covers some part of the set, behaves as:
\begin{equation}
\label{probabox}
P(\varepsilon) \sim \varepsilon^{1-d_C} \; .
\end{equation}
This probability also corresponds to the probability
that a hole (or a void) of $\cS_D$ around some arbitrary point $t_0$ has a size smaller than $\varepsilon$.
It results that the probability density function of $\tau$, the hole size of $\cS_D$ around some given position, behaves as
(when $\tau \ll T$):
\begin{equation}
\label{ptau}
p(\tau) \sim  \tau^{-d_C} \; .
\end{equation}

In fact, the exact shape of this probability law can be computed 
from the results of Ref. \cite{Fitz85}.
In this paper it is advocated that the set $\cS_D$  corresponds to 
the closure of the image of non-decreasing Levy process (i.e. a ``subordinator'').
If $\mu(t) = \mu([0,t])$ is a measure uniformely spread over the set $\cS_D$
(see Sec. \ref{sec:meas} for the definition of such a measure), this means that $t(\mu)$
is a non decreasing Levy process. Within this approach,
the void size statistical properties can be recovered from the jump size statistics of 
this subordinator. The relationship between the Levy process $t(\mu)$ and the set $\cS_D$ is made explicit by relating its Levy measure to the density $\Lambda(s)$ defined in Eq. \eqref{intensity}.

Let us define $G(s) = \int_s^\infty \Lambda(s) ds$.
From the expression of $\Lambda(s)$ in Eq. \eqref{intensity}, one has simply 
\begin{equation}
\label{defG}
G(s) =  \frac{(1-D)}{s} H(T-s)
\end{equation}
where $H(t)$ represents the Heaviside function.

Let us recall that the Laplace exponent  (or  cumulant generating
function) $\phi(\alpha)$ associated with the subordinator $t(\mu)$ is such that, for $\alpha > 0$,
$$ 
  \EE{e^{-\alpha t(\mu)}} = e^{-\mu \phi(\alpha)} \; .
$$
This function is related to the Levy measure $\nu(z)$ 
associated with the process as (see e.g. \cite{Fel71}):
\begin{equation}
\label{defphi}
  \phi(\alpha) = \int_0^\infty (1-e^{-\alpha z}) \nu(z) dz \; .
\end{equation}
According to standard results from potential theory, it is proven in
\cite{Fitz85} that $\phi(\alpha)$
can be directly related to the Laplace transform of 
$e^{\int_t^T G(u) du}$:
\begin{equation}
\frac{1}{\phi(\alpha)} =  \int_0^\infty e^{-\alpha s} e^{\int_s^T G(u) du} ds 
\end{equation}
From expression \eqref{defG}, we then have:
\begin{equation}
  \frac{1}{\phi(\alpha)} = T^{1-D} \int_0^T e^{-\alpha s} s^{D-1} ds + \int_T^\infty e^{-\alpha s} ds 
\end{equation}
and thus $\phi(\alpha)$ can be expressed in terms of the incomplete Gamma function $\gamma(z,t) = \int_0^t e^{-u} u^{z-1} du$ 
as:  
\begin{equation}
\label{phi_expr}
 \phi(\alpha) = \frac{\alpha}{T^{1-D} \alpha^{1-D} \gamma(D,\alpha T)+e^{-\alpha T}}
\end{equation}

Let $F(\tau) = \int_\tau^{\infty} \nu(z) dz$. Its Laplace transform can directly be obtained 
from Eq. \eqref{defphi} and \eqref{phi_expr} as
\begin{equation}
\label{laplaceF}
\int_0^{\infty} \! \! \! e^{-\alpha \tau} F(\tau) d\tau = \frac{\phi(\alpha)}{\alpha} =  \frac{1}{T^{1-D} \alpha^{1-D} \gamma(D,\alpha T)+e^{-\alpha T}}
\end{equation}

Let us recall that $N([\mu,\mu+w],\tau)$, the number of jumps of a Levy process 
of size greater than $\tau$ (and therefore the number 
of voids of size greater than $\tau$ in our problem) in some interval $[\mu,\mu+w]$, is a Poisson random 
variable of intensity $w F(\tau)$.
In that respect, since the average number of voids of size greater than $\tau$ is proportional 
to $F(\tau)$, $F(\tau)/F(\Delta)$ can be interpreted as the probability that a void
chosen at random among all voids of size greater than $\Delta$, has a size larger than $\tau$
\footnote{Notice that because $\nu(z)$ may not be integrable it is not rigorously speaking
a probability density function.}. This function will be denoted as $F_\Delta(\tau)$. 

Notice that selecting at random a void (over all the voids of size greater than some cut-off scale $\Delta$ in some time window) and considering the void around some arbitrary point $t_0$ do not provide the same probability
distributions as a function of the void size $\tau$: there must be an additional factor $\tau$ in the latter distribution in order to account for the fact that a void
of size $\tau$ ``contains" $\tau$ points and should be counted $\tau$ times.
This result can be directly proven using
Lemma 3 in \cite{Fitz85}, according to which 
the void size probability density $p(\tau)$ around some given point
can be explicitely written as proportional to both $\nu(\tau)$ and
the void size $\tau$: 
\begin{equation}
\label{vpdf}
 p(\tau) = \frac{\tau \nu(\tau)}{\int_0^\infty F(s) ds}
\end{equation}
However this distribution is hard to estimate from empirical data as compared to $F_\Delta(z)$ than can be directly estimated using a simple rank ordering of the observed void sizes (see Fig. \ref{fig:void}).

The analytical expression \eqref{laplaceF} does not allow us to obtain a closed-form expression
of $F(\tau)$, the void size distribution.
One can however obtain an estimation over some given range using 
numerical methods to invert the Laplace transform such as e.g. a Fourier method 
(see section \ref{sec:examples}).
Thanks to Tauberian like theorems \cite{Fel71,TailLaplace} one can also obtain
its asymptotic behavior in the limits $\tau \to \infty$ or $\tau \to 0$.
In appendix \ref{App:void_asympt}, we show that:
\begin{equation}
\label{Fasympt}
F(\tau) \sim \left\{
\begin{array}{ll}
  T^{-1} \left(\dfrac{\tau}{T}\right)^{-D} & \; \; \mbox{when} \; \;   \tau \to 0 \\ 
  T^{-1} \exp\left(-\dfrac{\alpha_D \tau}{T}\right) &  \; \;  \mbox{when} \; \;  \tau \to \infty 
\end{array}
\right.
\end{equation}
where $-\alpha_D$ is the largest real part of solutions of Eq. \eqref{pole_eq}. As illustrated in Fig. \ref{fig:alpha_D},
it goes from $0$ to  $\infty$ as $D$ varies from $0$ to $1$.
Note that this equation precisely means that the box-counting dimension of the set $\cS_D$ is $D$, i.e.,:
\begin{equation}
\label{eq:dimension}
  d_C = D
\end{equation}
(this directly results from \eqref{ptau} and Eqs (\ref{Fasympt},\ref{vpdf})).
Let us also remark that the behavior of $F(\tau)$ for large $\tau$ shows that the characteristic void size is 
not necessarily $T$ but $\frac{T}{\alpha_D}$ that can be very large (resp. very small) as respect to the integral scale $T$ 
when $D$ is close to 0 (resp. close to 1). Thus the ``effective integral'' scale also depends on the dimension $D$.

At this stage let us notice that the construction 
we propose involves a large cut-off scale $T$ above which
the fractal scaling no longer holds.  On a general ground, it is easy to show that if one supposes
that $\cS_D$ is a stationary self-similar random Cantor 
set of dimension $D$, then it necessarily involves
an integral scale $T$. 
Indeed, if one considers some time window ${\cal L}$ of size $L$, then, from the self-similarity of $\cS_D$, 
the (mean) number of boxes of size $\varepsilon$ necessary to cover $\cS_D \bigcap {\cal L}$ is
\begin{equation}
\label{scalingCantor}
N(\varepsilon,L) = K(L) \; \varepsilon^{-D} \; .
\end{equation}
From the stationarity of $\cS_D$, one must have $K(2L) = 2 K(L)$
and therefore
$$
  K(L) = K_0 L \; .
$$ 
But since $N(\varepsilon,L) \leq L \varepsilon^{-1}$ 
it results that the scaling \eqref{scalingCantor}
can hold only if 
$$
 \varepsilon \leq K_0^{\frac{-1}{1-D}}
$$
which shows that there necessarily exists an integral 
scale $T = K_0^{\frac{-1}{1-D}}$ setting an upper limit of the fractal scaling range. This definition allows one to rewrite Eq. \eqref{scalingCantor} for $\varepsilon \leq T$ as:
$$
  N(\varepsilon,L) = \frac{L}{T} \left( \frac{\varepsilon}{T}\right)^{-D}
$$

 


Instead of studying directly $\cS_D$, one can also consider
a measure $\mu$ which is supported by this set. In the next
section we introduce a multifratal class of such measures.
Some of the properties of the set $\cS_D$ will be recovered in the degenerated case when the multifractality
vanishes and $\mu$ is homogeneous over $\cS_D$.

\section{Continuous random cascades supported by a Cantor set}
\label{sec:meas}
\subsection{Definition and convergence}
Let us reconsider the previous construction with the goal of building 
a continuous cascade measure supported by $\cS_D$ and by this means, extending the construction of Bacry-Muzy. 
Let us define, as in ref. \cite{MuBa02,BaMu03}, 
\begin{equation}
\label{def-omega}
 \omega_\ell(t) = m(\cC_\ell(t)) = \int_{\cC_\ell(t)} dm(v,s)
\end{equation}
where $dm(t,s)$ is an infinitely divisible random measure (e.g. a normal or Poisson compound law)
and $\cC_\ell(t)$ is the cone set centered at $t$ as defined in section \ref{sec:casc} or by Eq. \eqref{def-cone} in Appendix \ref{cone_app}.

If $\psi(s)$ is the cumulant generating function as given by the celebrated 
Levy-Khintchine formula, then one has
(see Appendix \ref{cone_app})
\begin{equation}
\label{defpsi}
 \EE{e^{p \omega_\ell}} = e^{\psi(p) S(\cC_\ell)}
\end{equation}
where $S(\cC_\ell)$ is the ``area'' of the set $\cC_\ell$, i.e.,
\begin{equation}
  S(\cC_\ell) = \rho_\ell(0) = \int_{\cC_\ell}  s^{-2} dt ds = 1 + \ln (\frac{T}{\ell}) \; .
\end{equation}
where the function $\rho_\ell(t)$ is defined in Eq. \eqref{rhoexact} of Appendix \ref{cone_app}.
If $0 < D \leq 1$, one can always choose $\psi(p)$ such that 
\begin{equation}
\label{condpsi}
  \psi(1) = 1-D
\end{equation}
meaning that
\begin{equation}
\label{condpsi1}
 \EE{e^{\omega_\ell}} = (eT)^{1-D} \ell^{D-1}  \; .
\end{equation}
Let us now introduce, as in previous section, a Poisson measure $dp(t,s)$
homogeneous in time and of intensity $(1-D) s^{-2}$. 
One denotes as $(t_i,s_i)$ the events associated with this process and
\begin{equation}
\label{defgamma}
\gamma_\ell(t) = \int_{\cC_\ell(t)} dp(v,s) = \sum_{(t_i,s_i) \in \cC_\ell(t)} 1
\end{equation}
the number of events in the domain $\cC_\ell(t)$. Let us remark that $\gamma_\ell(t)$ represents
precisely the Poisson random variable of intensity $(1-D) \rho_\ell(0)$ that is the number of events $(t_i,s_i)$ in the $(t,s)$ plane (as introduced in sec. \ref{cutouts}) such that $s_i \geq \ell$ and $t \in \iI_{t_i,s_i}$.

One can then define a multifractal random measure as the limit when $\ell \to 0$ 
of the density:
\begin{eqnarray}
\label{defmu1}
 \frac{d\mu_\ell(t)}{dt} & = &  \; e^{\omega_\ell(t)} \prod_{s_i \geq \ell} \bI_{\iI_{t_i,s_i}}(0,t) \\
 \label{defmu2}
              & = &  \; e^{\omega_\ell(t)} \delta_{\gamma_\ell(t)}
\end{eqnarray}
where $\delta_n$, is the discrete delta function that is equal to $1$ if $n=0$ and 0 otherwise
(recall that $\gamma_\ell(t)$ is a Poisson random variable and, in that respect, takes integer values).
We show in Appendix \ref{conv_app} that $d\mu_\ell(t)$ weakly converges in the $L^2$ sense, when $\ell \to 0$,
towards a non-degenerated measure $d \mu$ provided $\psi(2) < 2-D$ \footnote{Along the same line 
	as in refs \cite{BaMan02,BaMu03}, it should be possible to use a martingale argument 
	and prove an almost sure version of the convergence with a weaker assumption but the $L^2$ version is sufficient
	for the purpose of this paper.}.
In the sequel $\mu(t)$ will stand for the limit measure of the interval $[0,t]$, i.e., 
$\mu(t)= \mu([0,t])=\lim_{\ell \to 0} \int_0^t d\mu_\ell(u)$, where the limit has to be interpreted in the mean-square sense.

Let us remark since, for all $t$, $\gamma_\ell(t)$ is a Poisson random variable of intensity $\lambda_\ell = (1-D) \rho_\ell(0)$, then  $\delta_{\gamma_\ell(t)}$ is simply a Bernoulli random variable of probability $e^{-\lambda_\ell}$.
In that respect, as in the random version of the $\beta$-model, one has:
\begin{equation}
\label{mu1p}
 d\mu_\ell(t) = \left\{ 
 \begin{array}{ll}
 0 & \; \; \mbox{with probability} \; \;   1-e^{-\lambda_\ell} \\ 
 e^{\omega_\ell(t)} dt & \; \;  \mbox{with probability} \; \;  e^{-\lambda_\ell} \; .
 \end{array}
 \right.
\end{equation}
This corresponds exactly to the way Schmitt defined his model of ``continuous
multifractal model with zero values'' (Eq. (31) of ref. \cite{Schmitt14}).
However, Eq. \eqref{mu1p} simply refers to the one dimensional marginal law of $d\mu_\ell(t)$ and is not equivalent to Eq. \eqref{defmu2} that defines the full process.
In order to illustrate that point, one can just remark that, according
to Eq. \eqref{f2} of appendix \ref{conv_app}, the Bernoulli variables $\delta_{\gamma_\ell(t)}$ have non trivial (power-law ) time correlations. 

\subsection{Stochastic self-similarity and scaling properties}
\label{sec:sss}
Let us now study the scaling properties of $\mu$ and show that it is a (stochastic) self-similar measure
that satisfies, for $s < 1$:
\begin{equation}
\label{ss-mu}
 \mu(s t) \stackrel{law}{=} s e^{\Omega_s} \delta_{\Gamma_s} \mu(t)
\end{equation}
where $\stackrel{law}{=}$ means an equality of all finite dimensional distributions.
$\Omega_s$ and $\Gamma_s$ are two random variables independent each other and independent 
of $\mu(t)$. Their law is respectively the same law as $\omega_{s T}$ and $\gamma_{s T}$.
The proof of this result 
relies on a variant of Lemma 1 of Ref. \cite{BaMu03} from which one can show that,
for $0 < s <1$ and for $u$ varying over an interval of size $T$:
\begin{eqnarray}
\label{ss-1}
   \omega_{s \ell}(s u) & \stackrel{law}{=} & \Omega_s + \omega_\ell(u) \\
   \label{ss-2}
   \gamma_{s \ell}(\lambda u) & \stackrel{law}{=} & \Gamma_s + \gamma_\ell(u)
\end{eqnarray}
where $\Omega_s$, $\Gamma_s$ are two independent random variables of the same
law as respectively $\omega_{s T}$ and $\gamma_{s T}$.
These stochastic equalities are proven in Appendix \ref{ss_app}. 
Thanks to the definition \eqref{defmu2}, one thus have:
\begin{eqnarray*}
 \mu_{\ell}(s t) & \stackrel{law}{=} & s e^{\Omega_s} \int_0^t e^{\omega_{\frac{\ell}{s}}(u)} \delta_{\gamma_{\frac{\ell}{s}}(u)+\Gamma_s} \; du  \\
 & \stackrel{law}{=} & s e^{\Omega_s} \delta_{\Gamma_s} \int_0^t e^{\omega_{\frac{\ell}{s}}(u)} 
 \delta_{\gamma_{\frac{\ell}{s}}(u)} \; du 
\end{eqnarray*}
where we used the fact that, for $n_1,n_2 \in \mathbb{N}$,  $\delta_{n_1+n_2} = \delta_{n_1} \delta_{n_2}$. 
Taking the (m.s)  limit $\ell \to 0$ of both sides leads to Eq. \eqref{ss-mu}.

Eq. \eqref{ss-mu} can be used to compute the scaling properties of the moments of measure 
(provided these moment exist). One defines the spectrum of scaling exponents $\zeta_q$ from 
the scaling behavior of the moment of order $q$ (when it exists):
\begin{equation}
\label{defzeta}
  M_q(t) \stackrel{def}{=} \EE{\mu(t)^q} \operatornamewithlimits{\sim}_{t \to 0} t^{\zeta_q}\; ,
\end{equation}
where $t \to 0$ means $t \ll T$.
This asymptotic behavior can be replaced by an exact power-law behavior
in the case of a self-similar processes.
Let $q > 0$ and let us consider $t \leq T$.  Since $\delta_n^q = \delta_n$, from
the self-similarity equation above, one has
\begin{equation}
 M_q(t) = T^{-q} t^q \EE{e^{q \Omega_{T^{-1}t}}} \EE{\delta_{\Gamma_{T^{-1}t}}} M_q(T) \; .
\end{equation}
Using Eq. \eqref{defpsi} and the fact that $\Gamma_s$ is a Poisson
random variable of intensity $(D-1)\ln(s)$, we thus obtain:
\begin{equation}
 M_q(t) = K_q t^{q-\psi(q)+1-D}
\end{equation}
with 
\begin{equation}
  K_q = T^{D-q-1+\psi(q)} M_q(T) \; .
\end{equation}
It follows that the multifractal spectrum of $\mu$ reads:
\begin{equation}
\label{eq:zeta_q}
 \zeta_q = q+1-D-\psi(q) \; .
\end{equation}
Let us remark that one recovers the result of the previous section about the box dimension of 
the Cantor set $\cS$ that supports the measure
$\mu$. Because $\mu(t)$ has stationary increments, in the limit $q \to 0$, $M_q(\varepsilon)$ can be interpreted as the probability that $\mu(\varepsilon)= \mu([t,t+\varepsilon]) > 0$, i.e., that a box of size $\varepsilon$ covers a part of $\cS$. From $M_0(\varepsilon) \sim \varepsilon^{1-D}$, one finds that $\cS$ is of fractal 
dimension $D$.

Let us also notice that, as in the case $D=1$ \cite{BaMu03}, 
the probability density function of $\mu(t)$ can have a fat tail. Indeed,
since if $q>1$, $M_q(t) > 2 M_q(t/2) = 2^{1-\zeta_q} M_q(t)$, one thus sees that,
if $\zeta_q < 1$ for $q>1$, then necessarily $M_q(t) = \infty$ meaning that the probability 
density function of $\mu(t)$ has an algebraic tail.

\subsection{Multifractal formalism}
\label{sec:mf}
Let us say few words about the multifractal formalism and some related issues such
as the scaling behavior of partition functions.
The multifractal formalism has been introduced in
early eighties by Parisi and Frisch (see e.g. \cite{ParFri85,Hasley86}) in order to interpret 
the multiscaling behavior of $M_q(t)$  
in terms of pointwise regularity properties of the paths of the process $\mu(t)$.
One defines the local
H\"older exponent $\alpha(t_0)$ at point (or time) $t_0$ as
\begin{equation}
\mu(t_0+\tau)-\mu(t_0) = \mu([t_0,t_0+\tau]) \operatornamewithlimits{\sim}_{\tau \rightarrow
	0} \tau^{\alpha(t_0)} \; .
\end{equation}
The singularity spectrum $f(\alpha)$ is then defined  as the fractal
(Hausdorff) dimension of the iso-singularity sets:
\begin{equation}
f^\star(\alpha) = dim_H \{ t, \alpha(t) = \alpha \} \; .
\end{equation}
Roughly speaking, this equation means that at scale $\tau \ll T$, the number of
points where $\mu([t_0,t_0+\tau]) \sim \tau^\alpha$ is 
\begin{equation}
N(\tau,\alpha) \sim \tau^{-f(\alpha)} \; .
\end{equation}
The {\em multifractal formalism} states that $f(\alpha)$ and $\zeta_q$ (as defined
in \eqref{defzeta})  are basically Legendre transforms to each other:
\begin{eqnarray*}
	f(\alpha) & = & 1+\min_q(q\alpha-\zeta_q) \\
	\zeta_q & = & 1 + \min_\alpha(q\alpha-f(\alpha)) \; .
\end{eqnarray*} 
The validity of the multifractal formalism is not straightforward to establish
but it is likely to hold for our construction since it enters in the general framework 
proposed in Ref. \cite{BaMan04}. This technical point is out of the scope of the paper and 
will be considered in a future work.

Let us notice that if $\zeta_q$ is concave and smooth, the previous Legendre transform
can be rewritten as
\begin{eqnarray}
\label{yes}
f(\alpha) &= &1 + q \alpha -\zeta_q \\
\alpha & = & \frac{d \zeta_q}{dq} 
\end{eqnarray}
If one denotes $\zeta_q^{(D)}$ the spectrum associated with the dimension $D$ in Eq. \eqref{eq:zeta_q}
(all other parameters remaining unchanged) and $f^{(D)}(\alpha)$ the associated singularity spectrum,
then from \eqref{yes} we have:
\begin{equation}
\label{falpha_D}
  f^{(D)}(\alpha) = f^{(1)}(\alpha)-(1-D),
\end{equation}
meaning that going from standard continuous cascades ($D=1$) to a version of dimension $D$
simply amounts to lower the singularity spectrum by $1-D$.

In order to estimate $\zeta_q$, one commonly uses the so-called partition function that can be defined,
for a sample of length $L$ as:
\begin{equation}
\label{pf}
 Z(q,\tau) = \int_0^L \mu([t,t+\tau])^q dt
\end{equation}
This function can be considered as an estimator of $M_q(\tau)$ and therefore
one thus expects that $Z(q,\tau) \sim \tau^{\zeta_q}$. In fact this is not always true
and if one defined $\zeta_q^\star$ such that
\begin{equation}
\label{zeta_star}
 Z(q,\tau) \sim \tau^{\zeta_q^\star}
\end{equation}
it is well known \cite{Mol97,MuBaBaPo08} that (for $q \geq 0$)
\begin{equation}
\label{lin_effect}
\zeta_q^\star = \left\{
\begin{array}{ll}
  \zeta_q & \mbox{if} \; q \leq q_\star \\
   1+\alpha_\star q & \mbox{otherwise}  
\end{array}
\right.
\end{equation}

where $q_\star = f'(\alpha_\star)$ and $\alpha_\star$ corresponds to the smallest singularity where the singularity spectrum vanishes, i.e.:
\begin{equation}
\label{alpha_star}
\alpha_\star  =  \min\{\alpha,~f(\alpha) = 0\}. 
\end{equation}
It is important to point out that, experimentally, under usual conditions, only $\zeta_q^{\star}(q)$ (and not $\zeta_q$) can be estimated and therefore the values of $\zeta_q$ are ``hidden'' when $q > q_\star$. Since according to Eq. \eqref{falpha_D}, for $D<1$
the $f(\alpha)$ spectrum is shifted downward, the smaller the dimension $D$ is, the smaller the value $q_\star$ and thus 
the domain over which one can estimate $\zeta_q$ will be 
(see next section).

\begin{center}
	\begin{figure}[h]
		\includegraphics[width=0.8 \textwidth]{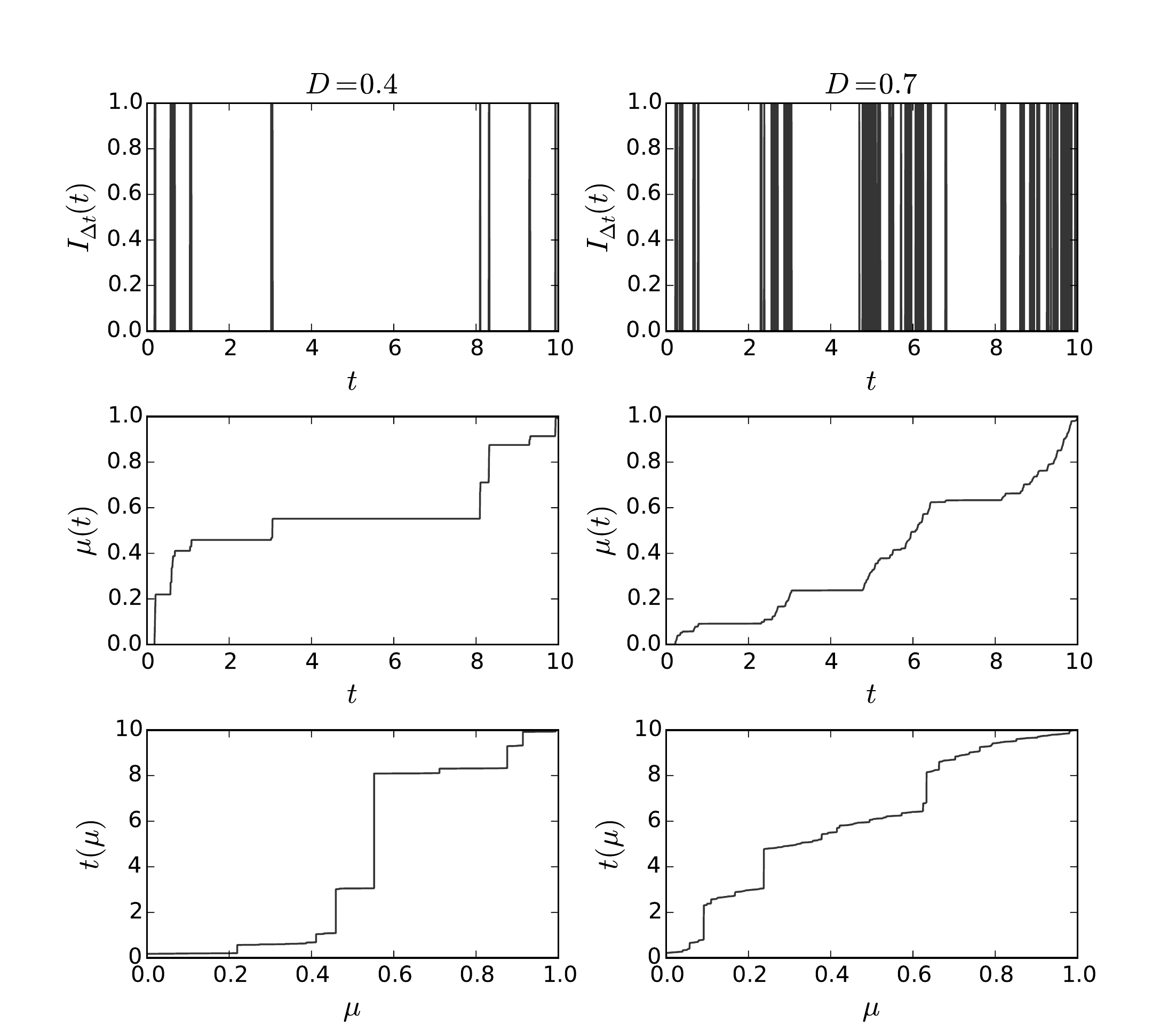}
		\caption{Examples of uniform measures spread on self-similar stationary Cantor sets $\cS_D$ of 
			dimension $D=0.4$ (left panels) and $D=0.7$ (right panels). In both cases the integral scale has been chosen 
			as $T=1$. On the top row are plotted 
			the indicator functions $I_{\Delta t}(t)$. On the middle row are plotted the measures $\mu(t)=\mu([0,t])$ while
			their inverse functions $t(\mu)$ are displayed in the plots of the third row. As advocated in section \ref{sec:void},
			$t(\mu)$ is a non-decreasing Levy process with Levy measure a given by Eq. \eqref{laplaceF}}
		\label{fig:cantor1}
	\end{figure}
\end{center}

\begin{center}
	\begin{figure}[h!]
		\includegraphics[width=0.8 \textwidth]{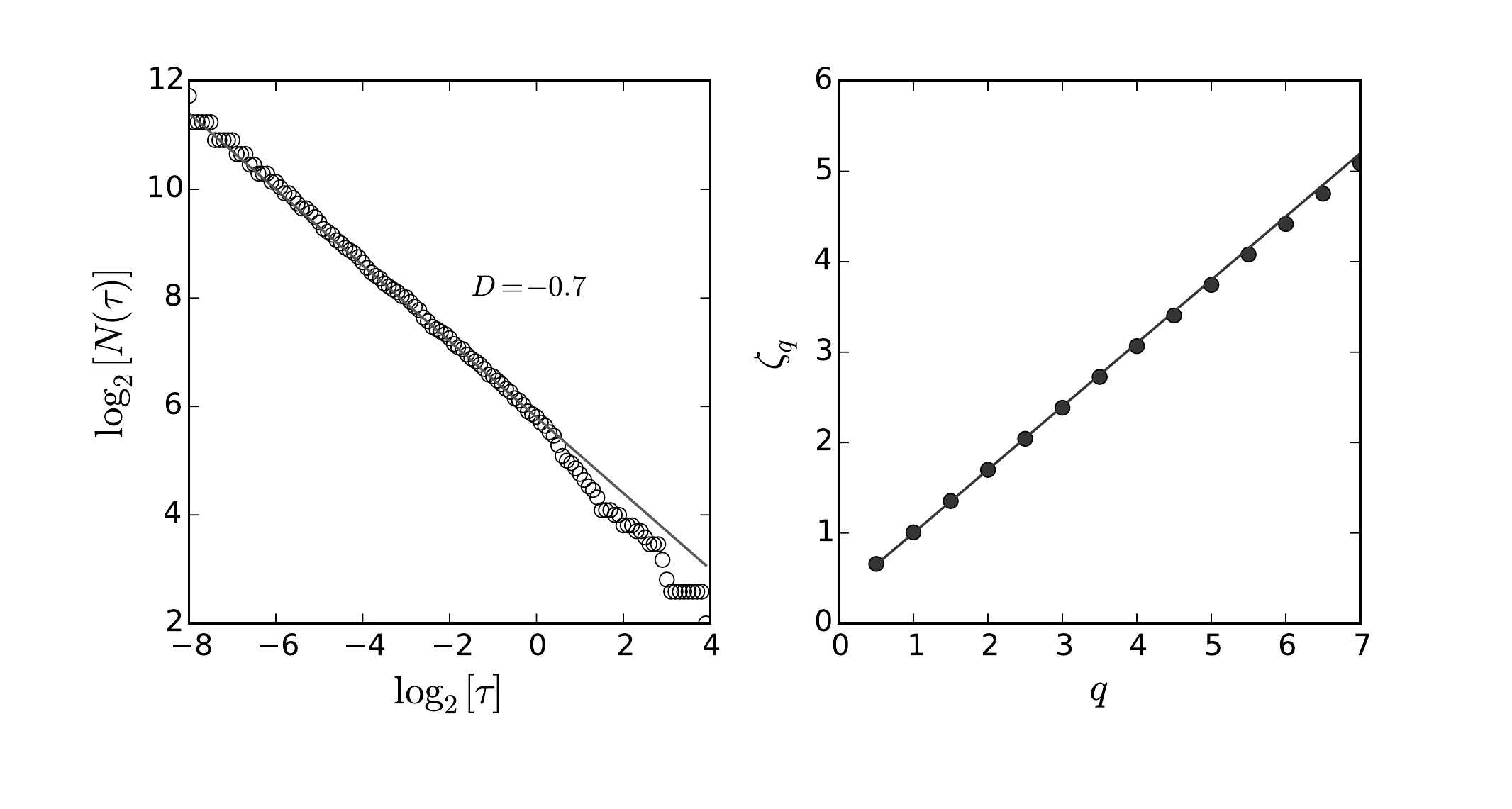}
		\caption{Scaling properties of the measure supported by a set of dimension $D=0.7$ displayed in Fig. \ref{fig:cantor1} 
			(middle right panel).
			Left panel: estimation of the box-counting dimension using WTMM (see text). We have displayed in log-log scale, the 
			WTMM number as a function of the scale $\tau$. The solid line corresponds to a line of slope -0.7. One can see that the scaling is no longer valid for scales greater than the integral scale $T=1$.
			Right panel: estimated $\zeta_q$ function obtained from the scaling behavior of empirical estimates of WTMM moments $M_q(\tau)$.
			The solid line represents the fit $\zeta_q = 0.7 q +0.3$ as predicted by Eq. \eqref{zetamono}.}
		\label{fig:scalingMono}
	\end{figure}
\end{center}

\begin{center}
	\begin{figure}[h!]
		\includegraphics[width=0.8 \textwidth]{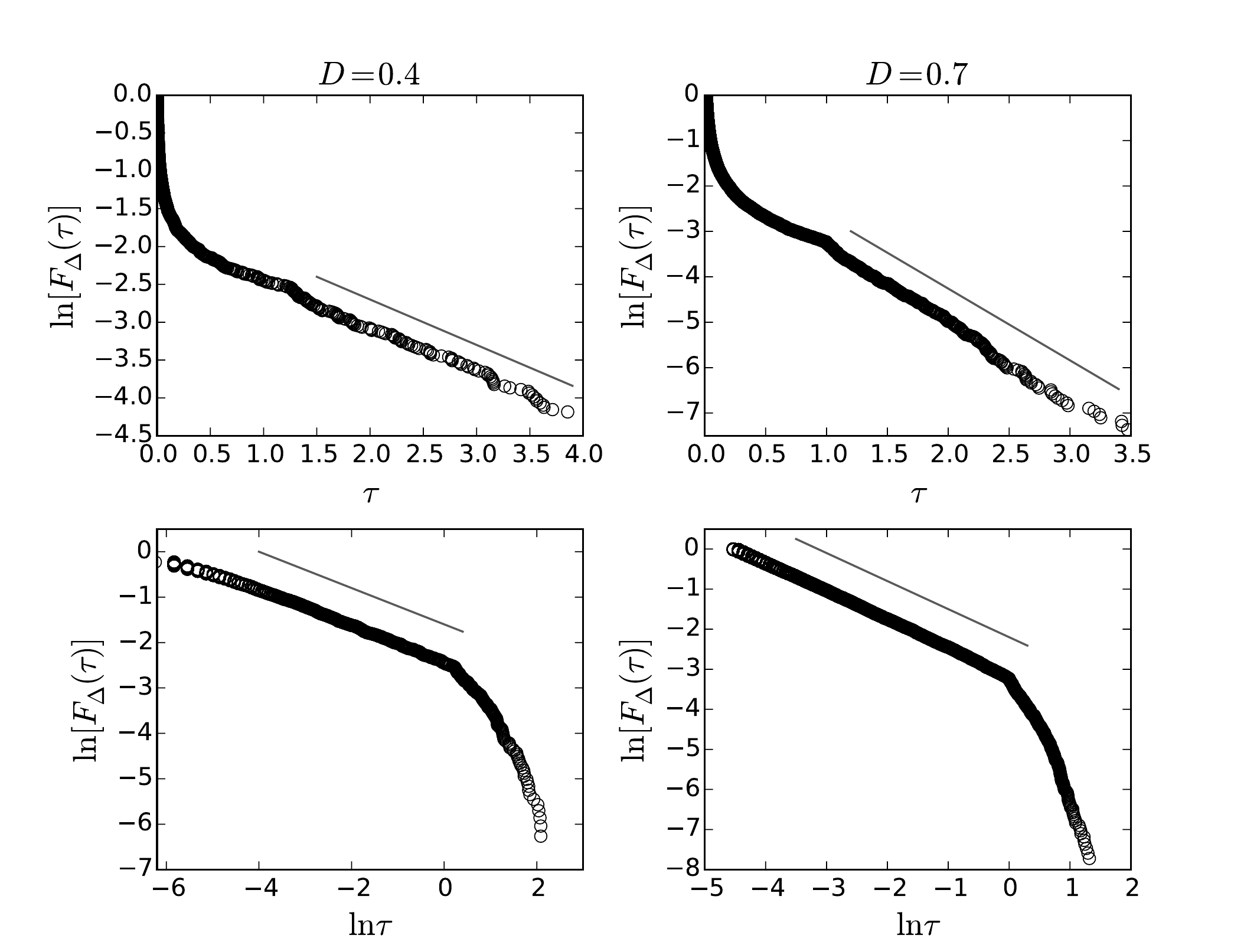}
		\caption{Empirical void size statistics of $\cS_D$ for $D=0.4$ (left panels) and $D=0.7$ (right panels). The distribution $F_\Delta(\tau)$ as defined in Sec. \ref{sec:void} is estimated from a rank ordering of the observed void sizes in simulated samples of $\cS_D$ of integral scale $T=1$, over an interval of width $40$ and at resolution $\Delta = 2^{-8}$. The small and large scale asymptotic regimes of Eq. \eqref{Fasympt} are empirically observed. In the top panels a semi-logarithmic representation shows the exponential tail. One can see that the theoretical prediction $-\alpha_D \tau$ provides a good fit (solid line). In the bottom plots a log-log scale shows that the regime when $\tau \to 0$ is a power-law $\tau^{-D}$
		(the solid lines represent the theoretical prediction).}
		\label{fig:void}
	\end{figure}
\end{center}

\begin{center}
	\begin{figure}[h]
		\includegraphics[width=0.6 \textwidth]{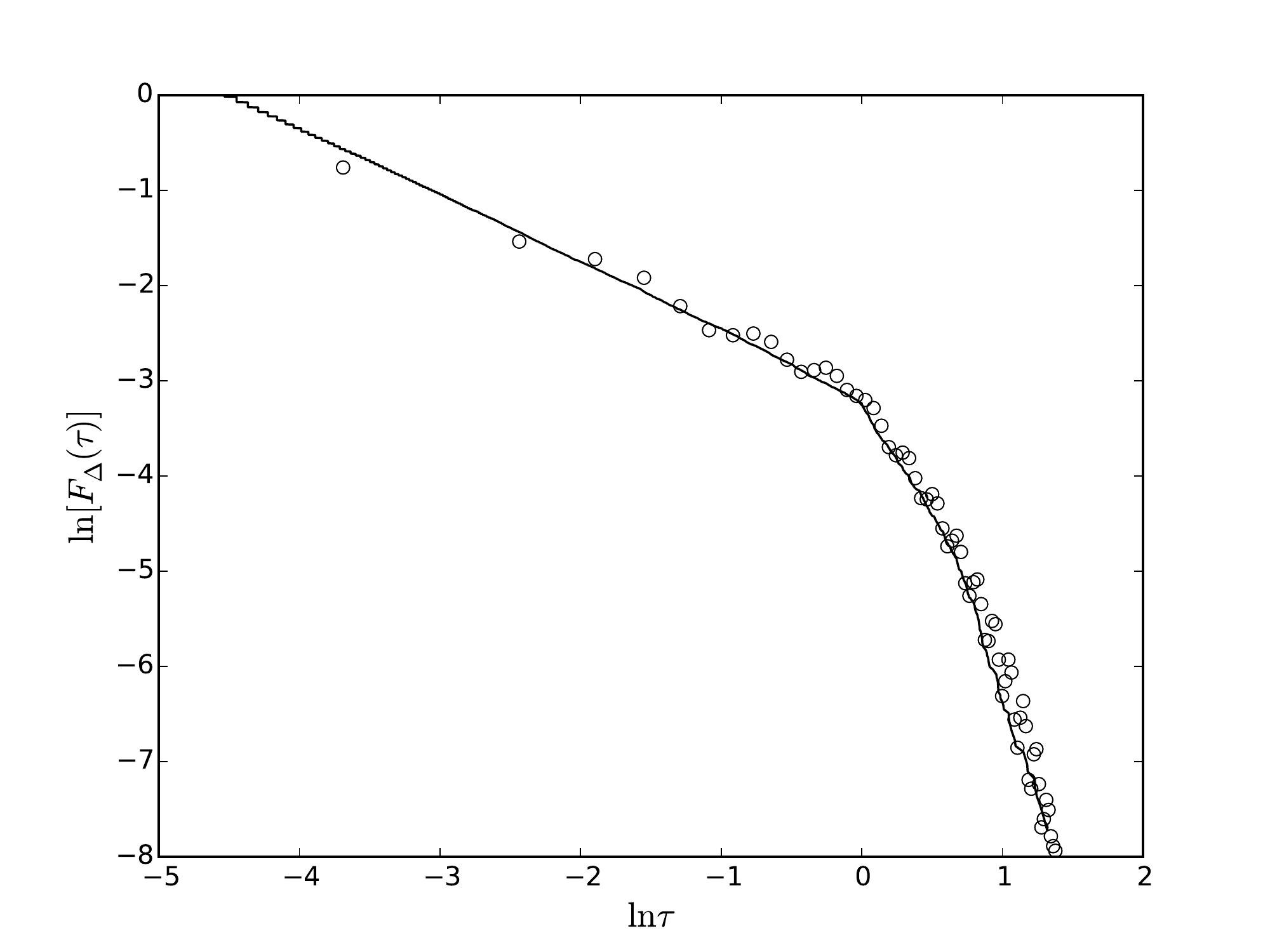}
		\caption{Log-log representation of empirical (continuous line) and theoretical ($\circ$) void distributions 
			$F(\tau)$ for the example $D=0.7$ of Fig. \ref{fig:void}. The ``theoretical'' values have been obtained using a simple numerical inversion of the Laplace transform \eqref{laplaceF} by a Fourier method.} 
		\label{fig:invlaplace}
	\end{figure}
\end{center}

\section{Simulations, numerical illustrations and application to high resolution rainfall data}
\label{sec:examples} 
Let us present various numerical simulations in order to 
illustrate previous considerations. A numerical method
to generate an infinitely divisible process $\omega_\ell(t)$
as defined in section \ref{sec:meas} (or in Appendix \ref{cone_app})
has been described e.g. in Refs \cite{MuBa02,ChaRieAbr05}.
Notice that in the case of a Gaussian $\omega_\ell(t)$, a simpler 
and faster FFT method can be directly used.
In order to simulate samples of $\delta_{\gamma_\ell(t)}$ in Eq. \eqref{defmu2}, one starts with a signal of constant value sampled at frequency $2 \ell$ and one discretises the logarithms of the scales $\ln(\ell) \leq v = \ln(s) \leq \ln(T)$ in small strips of width $\Delta v$ (typically one chooses $\Delta v$ in order to have around 100 strips between $\ell$ and $T$). 
One then generates in each strip a Poisson point 
process in time of intensity $(1-D) e^{-v} \Delta v$ as given by
Eq. \eqref{intensity}. For each of the obtained events $(t_i,s_i)$, we set to zero all the signal values in the interval $(t_i-s_i/2,t_i+s_i/2)$.
The examples considered in this section are built in that way.
\begin{center}
	\begin{figure}[h!]
		\includegraphics[width=0.7 \textwidth]{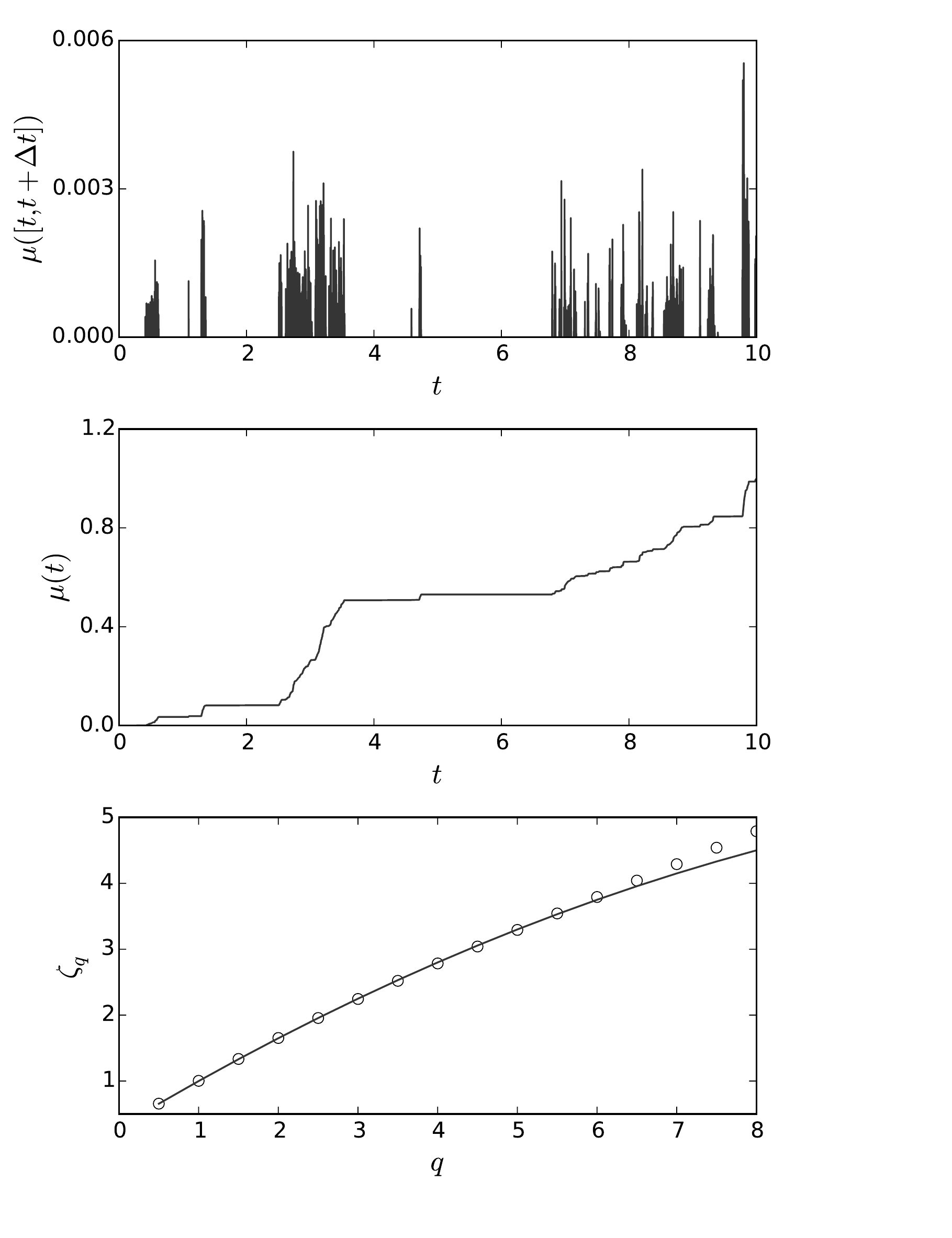}
		\caption{Example of a log-normal multifractal measure supported by a set of dimension $D=0.7$.
			In the top panel is represented a sample of the measure $\mu([t,t+\Delta t])$ where $\Delta t = 2^{-10}$.
			The integral scale is $T=1$, the fractal dimension is $D=0.7$ and the intermittency coefficient is $\lambda^2 = 0.05$.
			The associated function $\mu(t)$ is plotted in the middle panel. In the bottom panel are reported the results of the estimation
			of the multifractal scaling exponent $\zeta_q$ ($\circ$) as compared to the theoretical expression \eqref{zetaqln} (solid 
			line).} 
		\label{fig:cmrw}
	\end{figure}
\end{center}

\subsection{Monofractal case: an homogeneous measure on the Cantor $\cS_D$}

The simplest construction of $\mu(t)$ according to definition 
\eqref{defmu2} corresponds to a non random $\omega_\ell(u)$, and therefore, in order to satisfy the condition \eqref{condpsi} (or Eq. \eqref{condpsi1}), to the choice:
\begin{equation}
   \omega_\ell = (D-1)\ln \ell + \ln C \; .
\end{equation}
In this case the measure is homogeneously distributed on the Cantor set $\cS_D$ 
and the spectrum $\zeta_q$ is a linear function, 
(the process is a so-called ``monofractal process''):
\begin{equation}
\label{zetamono}
 \zeta_q = q+(D-1)(q-1) = 1-D+Dq \; .
\end{equation}
Two numerical samples of such a measure are plotted in top panel 
of Fig. \ref{fig:cantor1} corresponding respectively to $D=0.4$ (left panel)
and $D=0.7$ (right panel). We chose $T=1$ in both cases.
Note that since $\mu(t)$ is a singular continuous measure (it does not have a density) one cannot plot $d\mu(t)/dt$.
For the sake of clarity we have represented the function $I_{\Delta t}(t)$, where $\Delta t$ is a small time resolution and $I_{\Delta t}$ is such that 
$I_{\Delta t}(t) = 1$ if $\mu([t-\Delta t/2,t+\Delta t/2]) > 0$ and 
$I_{\Delta t}(t) = 0$ otherwise. This function can be seen 
as the finite scale resolution of the indicator function of the random
Cantor sets $\cS_D$. The functions $\mu(t)$ are plotted 
in the middle panel. Notice that $\mu(t)$ are singular but continuous 
time functions. This is not the case of their inverse functions $t(\mu)$
displayed in the bottom figures whose jumps correspond to the voids
of the Cantor sets. According to the results of section \ref{sec:void}, $t(\mu)$ are Levy processes associated with the Levy measure given by the derivative of $F(\tau)$ which Laplace transform corresponds to expression \eqref{laplaceF}.

In order to numerically check Eq. \eqref{zetamono}, one can direcly estimate
from the data the partition functions as defined in Eq. \eqref{pf}.
Alternatively one can use the so-called
Wavelet Transform Modulus Maxima (WTMM) 
method based on the scaling properties of wavelet coefficients
at modulus maxima \cite{MuzBacArn91,BacMuzArn93,MuzBacArn94}. 
In Fig. \ref{fig:scalingMono} are displayed the results of such an analysis
for the case $D=0.7$.
In the left panel we have plotted, in double logarithmic scale, the number
$N(\tau)$ of wavelet transform maxima as a function of the scale. This number
is proportional to the number of boxes of size $\tau$ necessary to cover 
the Cantor and therefore provides an estimation of the box-counting dimension $d_C$. In the left fig. \ref{fig:scalingMono}, one estimates a dimension very close to $D=0.7$ (corresponding to the slope of the solid line). The estimation
of the $\zeta_q$ spectrum from the scaling of higher order WTMM moments
is represented by $(\bullet)$ symbols in the right panel. One can check the excellent agreement with the expression \eqref{zetamono} corresponding to $D=0.7$ (solid line). 

Previous numerical simulations can also be used to verify the theoretical 
considerations of section \ref{sec:void}.
In Fig. \ref{fig:void} are plotted the estimated void distribution
for both $D=0.4$ and $D=0.7$ cases. $F_\Delta (\tau)$ (where $\Delta$ is the sampling period) is simply estimated as the normalized rank-ordered distribution of the observed hole sizes. In both semi-logarithmic plot
(top panels) and double-logarithmic plots (bottom panels) of the estimated 
distributions, one can see that the predictions of Eq. \eqref{Fasympt} represented by the solid lines are very well verified.
In the case $D=0.7$, one can check in Fig. \ref{fig:invlaplace} that the Laplace transform \eqref{laplaceF} accounts for the full shape 
of $F_\Delta(\tau)$: the ($\circ$) symbols represent the computed values 
of $F_\Delta(\tau)$ by a numerical inversion of the Laplace transform \eqref{laplaceF} (using a FFT method) while the solid line represents the distribution
estimated from simulation data.

\subsection{Multifractal examples}
The simplest example of a multifractal cascade is the 
log-normal case which corresponds to a measure $dm(t,s)$ that is a Wiener noise 
of variance $\lambda^2$
or, equivalently, to $\omega_\ell(u)$ that is a stationary Gaussian
process of covariance $\lambda^2 \rho_\ell(\tau)$ with $\rho_\ell$ as given by Eq. \eqref{rhoexact}. In that case, 
one has simply $\psi(p) = pm+\frac{\lambda^2}{2}p^2$ and the condition \eqref{condpsi}
leads to $m = 1-D -\frac{\lambda^2}{2}$. The multifractal spectrum is therefore
\begin{equation}
\label{zetaqln}
\zeta_q = 1-D+q(D+\frac{\lambda^2}{2})-\frac{\lambda^2}{2}q^2
\end{equation}
which extends the standard log-normal spectrum to arbitrary $D < 1$. 
The coefficient $\lambda^2 = -\zeta''(0)$ is called the intermittency coefficient
and quantifies the level of multifractality of the measure. In the limit $\lambda^2 \to 0$, one 
recovers the monofractal case \eqref{zetamono} and if $D=1$ one recovers the
MRW log-normal measure introduced in \cite{MuDeBa00,BaDeMu01}.
A sample of a log-normal cascade with $D=0.7$, $T=1$ and $\lambda^2 = 0.02$
is displayed in Fig. \ref{fig:cmrw} (top and middle panels).
As in the previous case, the fractal dimension $d_C$ and the 
$\zeta_q$ spectrum can be estimated from a WTMM scaling analysis.
These estimations from a single sample of length 128 $T$ at a resolution $\Delta t = 2^{-10}$ are displayed in the bottom panel of 
Fig. \ref{fig:cmrw} ($\circ$). The solid line represents the analytical 
expression \eqref{zetaqln} with $D = 0.7$ and $\lambda^2 = 0.05$.
One can see that the agreement is very good up to a value $q \simeq 5.5$.
For greater $q$ values, the estimated $\zeta_q$ spectrum appears to be a linear function of slope around $0.5$. This observation corresponds exactly
to Eq. \eqref{lin_effect} at the end of section \ref{sec:mf}.
Indeed, the scaling of the partition function (as estimated by the WTMM method
or by any other method) defines the spectrum $\zeta_q^\star$ that, according
to Eq. \eqref{lin_effect}, differs from $\zeta_q$ for $q > q_\star$.
In that domain, $\zeta_q^\star$ is a linear function of slope $\alpha_\star$ as given 
by Eq. \eqref{alpha_star}. From Eqs \eqref{yes} and \eqref{alpha_star} together
with expression \eqref{zetaqln}, one gets:
\begin{equation}
  q_\star = \sqrt{\frac{2D}{\lambda^2}} \; \mbox{and} \; \alpha_\star = \frac{\lambda^2}{2}+D-\sqrt{2D \lambda^2} \; .
\end{equation}
The values $D = 0.7$ and $\lambda^2 = 0.05$ leads to $q_\star \simeq 5.29$ and $\alpha_\star \simeq 0.46$ in excellent agreement with our 
numerical observations. Notice that, for a log-normal cascade, $q_\star \to 0$
when $D$ decreases or when $\lambda^2$ increases. This means that for high intermittent
cascades on a support of small dimension $D$, the range where one can observe a parabolic 
$\zeta_q$ can be very small.

Many other examples than the log-normal cascade can be considered.
For example, a Poisson compound 
corresponds to $\psi(p)= m p + K \int (e^{px}-1) F(dx)$
where $F(dx)$ is a probability distribution of some positive random variable.
If $W>0$ is a random variable such that the law of $\ln(W)$ is $F$, then the $\zeta_q$ spectrum
of the log-Poisson compound cascade can be written as:
\[
 \zeta_q = 1-D + mq + K (\EE{W^q}-1)
\]
where $m$ is such that $\zeta_1 = 1$. This spectrum generalizes to non integer dimensions the spectrum of Barral-Mandelbrot cascade model \cite{BaMan02}.
Let us also mention log-$\alpha$-stable random cascades 
that correspond to $\psi(p)=  \kappa (p^\alpha-p) + (1-D)p$ and therefore to 
\[
 \zeta_q =  1-D + q (D+\kappa) - \kappa q^{\alpha}
\]
which extends the log-normal spectrum to Levy indexes $0 < \alpha \leq 2$.
Such laws have been notably used in the context of turbulence and geophysics \cite{scherlev}.
Along the same line, other families of $\zeta_q$ spectra can be obtained
(e.g. log-Gamma, log-Hyperbolic,...) for alternative
choices of infinitely divisible generating function $\psi(p)$.
\begin{center}
	\begin{figure}[h!]
		\includegraphics[width=0.7 \textwidth]{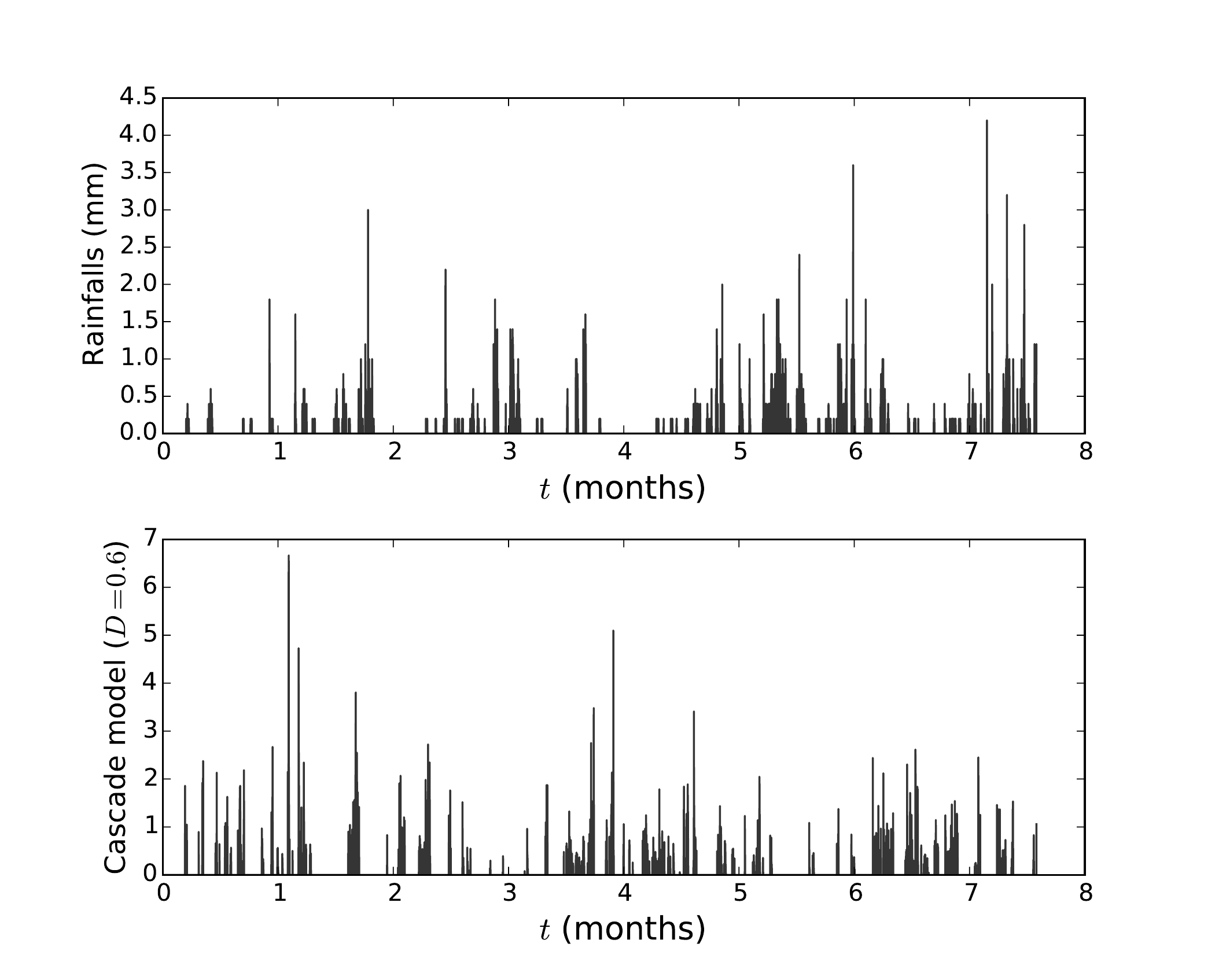}
		\caption{Sample of the rainfall data we used: on the top panel is displayed a sample of 6 mn frequency
			rainfall intensity at Anglet over a period of 8 months during 2006. The bottom figure represents a sample of the log-normal model lying over a set of dimension $D=0.6$, with intermittency coefficient 
			$\lambda^2 = 0.07$ and integral scale $T = 5$ days.}
		
		\label{fig:rain1}
	\end{figure}
\end{center}

\begin{center}
	\begin{figure}[h]
		\includegraphics[width=0.9 \textwidth]{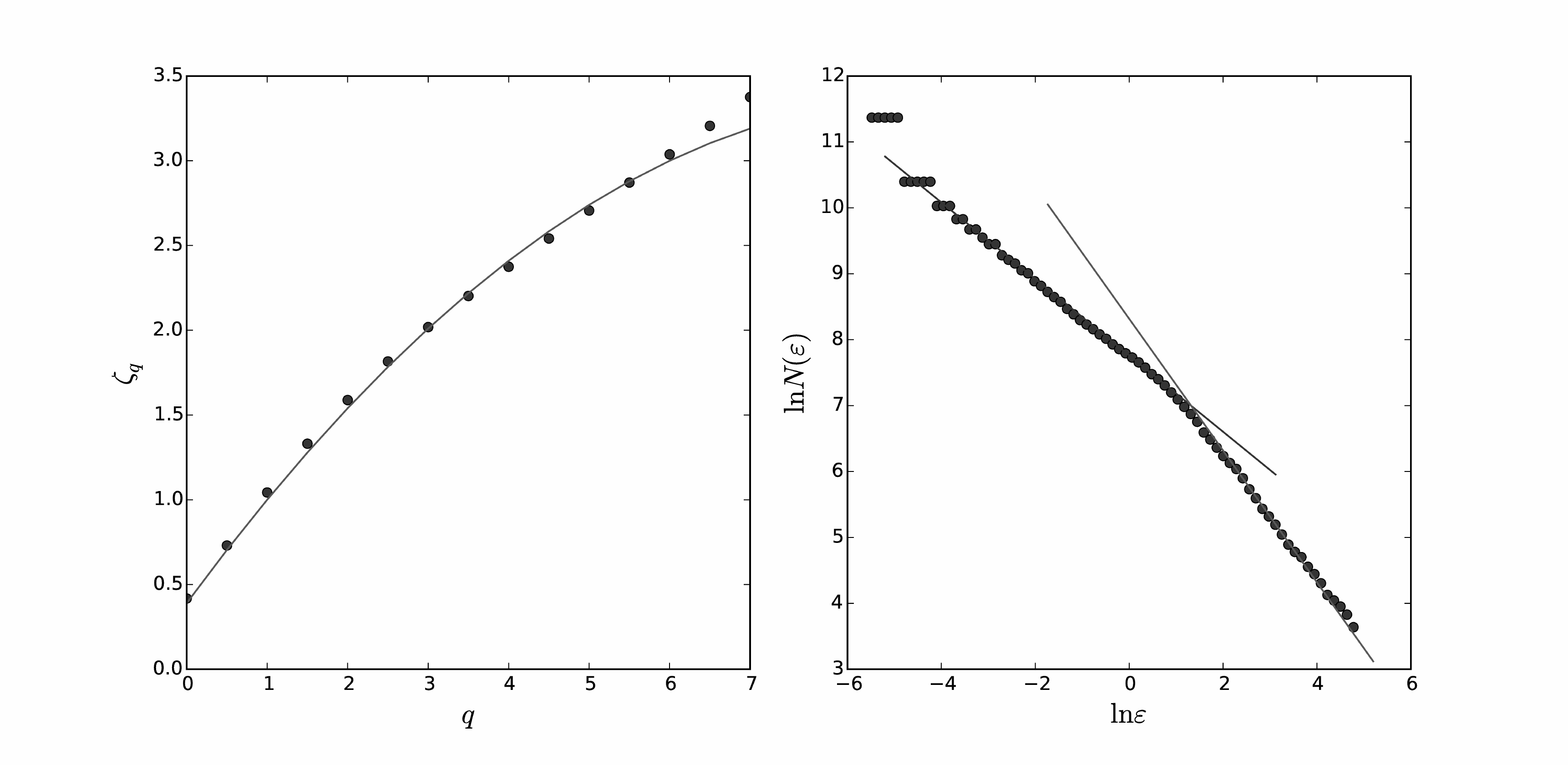}
		\caption{Multifractal analysis of the 6-min rainfall data by the mean of the WTMM method (see text). Left panel: the empirical $\zeta_q$ ($\bullet$) is fitted by a log-normal spectrum with $D = 0.6$ and $\lambda^2 = 0.07$ (solid line). Right Panel: the behavior of the number of boxes necessary to cover the rainy periods is plotted as a function of the box size $\varepsilon$ in double logarithmic scale. The so-estimated fractal  dimension (corresponding to the slope of this line) is $D \simeq 0.6$ (the fit is represented by the left solid line). One can observe a large scale $T$ above which $N(\varepsilon) \sim \varepsilon^{-1}$ (right solid line). This corresponds to the integral scale of the model which is estimated as $T \simeq 5$ days.}		
		\label{fig:rain2}
	\end{figure}
\end{center}

\begin{center}
	\begin{figure}[h]
		\includegraphics[width=0.7 \textwidth]{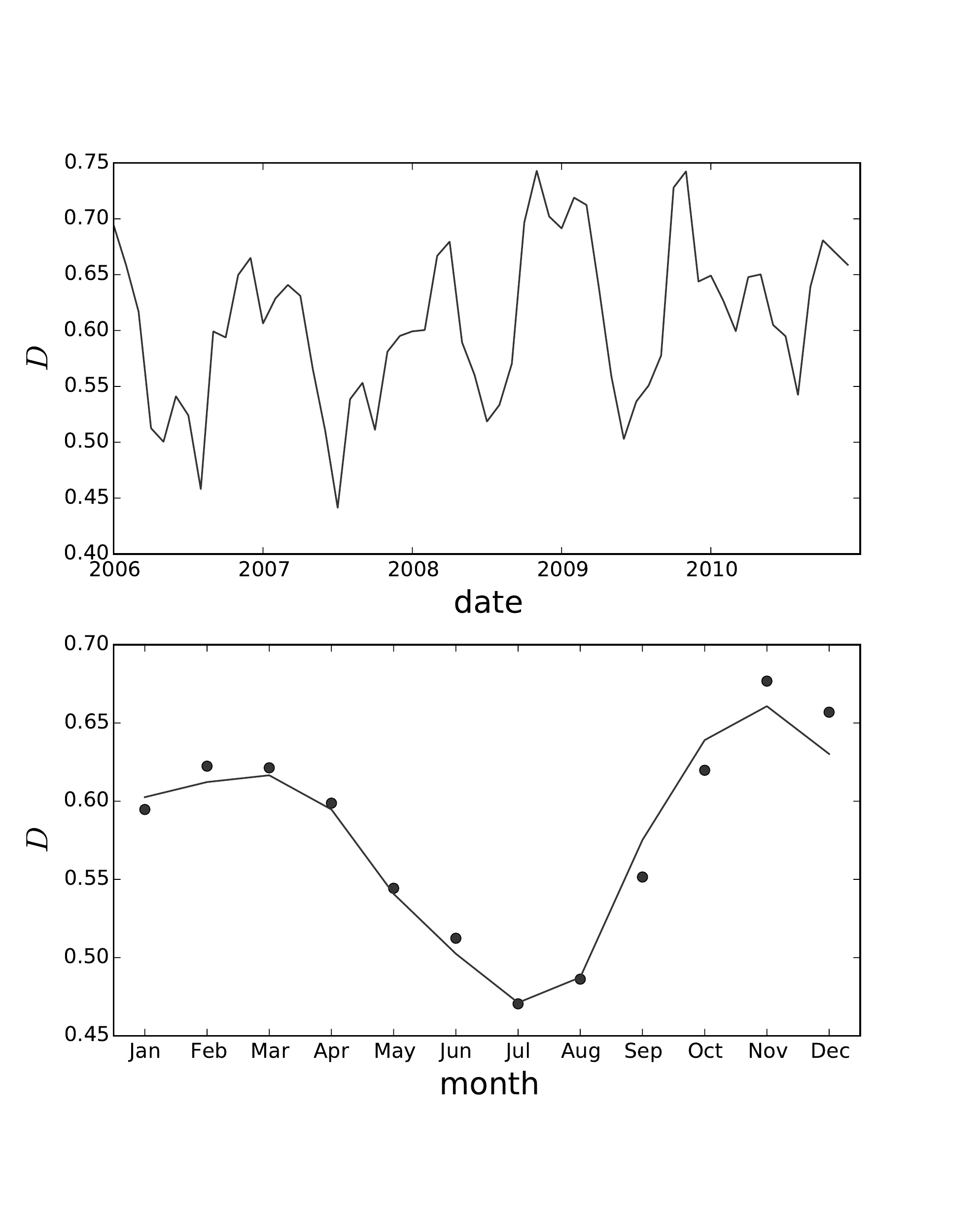}
		\caption{Seasonal variations of the fractal dimension of the set of rainy periods. Top panel: 
			$D$ is estimated on a two months sliding window from the set of 5 time series. One can observe that the fractal dimension decreases from values around $2/3$ during winter to a value smaller than $1/2$ in summer. Bottom panel: seasonal variation of the fractal dimension as obtain from the mean value over all series and all years. $(\bullet)$ symbols represent the estimations performed from the WTMM method while the solid line represents the estimation performed from the behavior of void size distribution around the origin (see text).}		
		\label{fig:rain3}
	\end{figure}							
\end{center}

\subsection{Application to high resolution rainfall data}
Precipitations result from a complex process that involves 
a large number of physical phenomena
from micro-physics inside the clouds to 
large scale meteorology. It is well known that the rainfall distribution displays a very high variability over a wide range of space and time scales \cite{LimaThesis}. Very much like turbulence, this intermittent behavior that can be hardly described from dynamical equations, has been successfully described within the framework of multifractal processes and random cascade models. This approach, pioneered by Schertzer and Lovejoy \cite{SL87}, Gupta and Waymire \cite{Gupta2} is still the subject of a very active research (see e.g \cite{LimaThesis,CeresettiThesis} and reference therein).

In this section, we simply want to illustrate the possible interest of the model introduced 
in this paper to describe the time variability of high resolution rainfall data 
\footnote{A study fully devoted to this topic with a comprehensive comparison of our model
to former approaches will be addressed in a forthcoming work.}. 
In fact the continuous cascade model described in section \ref{sec:mf} can be considered
as an extension of the models of refs. \cite{Over96,Schmitt98} where the framework of random cascades on fractal sets has been used 
to account for both rainfall intensity fluctuations and the occurrence of wet and dry events.
We have analyzed a set of 5 time series corresponding to high-resolution precipitations recorded at 5 different locations in the south west of France (see table \ref{tab1}). The series were provided by 
Meteo-France and have been recorded by the means of tipping-bucket rain gauge with a resolution of 6 min in time and 0.2 mm of equivalent rainfall depth. The measurements are available over 5 years from the begin of 2006 to the end 2010.
A sample of the rainfall series at Anglet corresponding to 8 months during 2006 
is displayed in the top panel of Fig. \ref{fig:rain1}. For comparison, a sample of a log-normal model of dimension $D=0.6$, integral scale $T = 5$ days and intermittency coefficient $\lambda^2 = 0.07$
is plotted in the bottom panel (see below for the choice of the parameters). One can see that, from a qualitative point of view, the model reproduces quite well both rainfall intensity and rainfall occurence fluctuations.
\begin{table}[ht]
	\begin{tabular}{|c|c|c|}
		\hline 1 & Aicirits-Camou-Suhast  & 43$^o$20'N  -  1$^o$01'W \\ 
		\hline 2 & Anglet   & 43$^o$28'N  -  1$^o$32'W   \\ 
		\hline 3 & Bustince-Iriberry     & 43$^o$10'N  -  1$^o$12'W  \\ 
		\hline 4 & Ciboure & 43$^o$24'N  -  1$^o$41'W \\ 
		\hline 5 & Orthez   & 43$^o$30'N  -  0$^o$45'W  \\ 
		\hline 
	\end{tabular}
	\caption{The locations of the 5 rainfall series}
	\label{tab1}
\end{table}

The estimated multifractal spectrum $\zeta_q$ obtained with the WTMM method over 
the whole 5 years period and by averaging the results of the 5 series is reported in the left
panel of Fig. \ref{fig:rain2} ($\bullet$). 
The fits of the power-law behavior of the partition function for all values 
of $q$ have been performed between the scales $\varepsilon = 24$ min and $\varepsilon = 2$ days. 
It is well known that, from a statistical point 
of view, it is extremely hard to determine the nature of the infinitely
divisible law of some cascade sample from empirical scaling of the partition 
function. As far as rainfalls are concerned, one can find very different fits
in the literature from log-normal \cite{Ven2002,Over96} to more general 
log-Levy cascades \cite{SL87,LimaThesis}. 
The solid line in the left panel of Fig. \ref{fig:rain2} 
represents a fit of $\zeta_q$ by 
a log-normal model (Eq. \eqref{zetaqln}) corresponding to $D=0.6$ 
and $\lambda^2 = 0.07$. We do not claim that a log-normal cascade provides
the best fit of the data but it allows one to reproduce the non-linearity
of the empirical spectrum quite well.  

In the right panel we have displayed, in log-log representation, the number of wavelet maxima $N(\varepsilon)$ at scale $\varepsilon$ (that 
is equivalent to the number of boxes of scale $\varepsilon$ necessary to cover the rainy periods) as the function of the scale $\varepsilon$. Very much like what we observed for the model in Fig. \ref{fig:scalingMono}, one clearly sees two regimes separated by a characteristic scale $T \simeq 5$ days: the small scale regime $\varepsilon \leq T$ where $N(\varepsilon) \sim \varepsilon^{-0.6}$ and the large scale regime where $N(\varepsilon) \sim \varepsilon^{-1}$. Such a measurement thus allows us to estimate both the dimension $D$ ($\simeq 0.6$) and the integral scale $T$ ($\simeq 5$ days) of the model.
Let us remark that the existence of a scaling range between few minutes and few days has been already pointed out in many studies devoted to rainfalls (see. e.g. \cite{Fabry96,CeresettiThesis}) and corresponds to time scales between 
turbulent motions and weather front systems. This scaling range has also been observed from the fluctuations of surface wind velocity \cite{MuBaPo10,BaiMu10}.
\begin{center}
	\begin{figure}[h]
		\includegraphics[width=0.9 \textwidth]{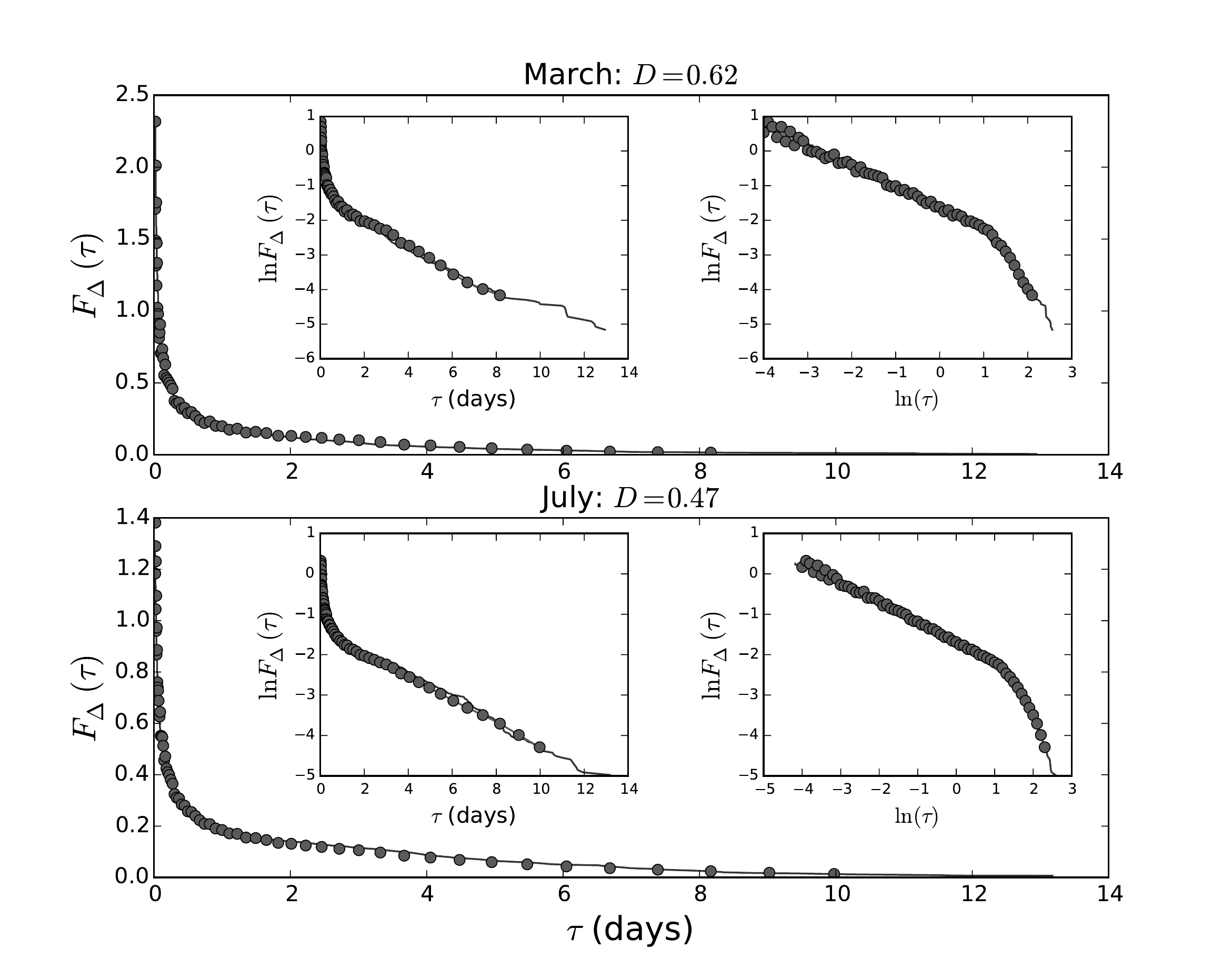}
		\caption{Modeling the distribution of dry period durations. In the top panel is plotted the estimated distribution of dry period durations from the 5 rainfall series in March (solid line) while in the botton panel is displayed the distribution estimated in July. In all figures, symbols ($\bullet$) represent the analytical predictions as obtained from the Laplace transform expression \eqref{laplaceF} with $T = 5$ days and the respective fractal dimensions $D=0.62$ (top panel) and $D=0.47$ (bottom  panel). In the insets, the same distributions are represented in semi-logarithmic (left insets) and double-logarithmic (right insets) scales. In all representations, one sees that the model predictions reproduce very well the empirical distributions.}		
		\label{fig:rain4}
	\end{figure}
\end{center}
The properties of the rainfall time series we just discussed
are average properties estimated over the whole year. 
In fact, it is well known  
that precipitations are a seasonal process so that one expects some 
variations of its fractal and multifractal properties through the seasons.
In order to check this feature, we have estimated the fractal dimension
of all series on a sliding window of width 2 months. The results we 
obtained are reported in Fig. \ref{fig:rain3}. In the top panel, the fractal dimension is plotted as a function of the date all along the 5 years 
period while in the bottom panel it is plotted 
as a function of the month (it is obtained by an average 
over the 5 years). One clearly sees that the fractal dimension $D$ varies 
from $D \simeq 0.47$ during the dry season in summer to $D \simeq 0.7$
in winter. Let us mention that comparable results 
have already been observed in former studies using hourly rainfall database (see e.g. \cite{LimaThesis}).
It thus appears that, within the framework introduced in this paper, the 
variation of the fractal dimension parameter $D$ can account 
for the variation of the occurrence likelihood of rainy periods through the seasons. In order to confirm this point and to check the model prediction
in a more precise way, one can compare the model void distribution as described 
in section \ref{sec:void} to the empirical distribution of dry period durations
as observed in the rainfall series. 
There have been few attempts in former studies to reproduce the likelihood of occurrence of wet and dry periods within the framework of random cascade models (see e.g. \cite{Schmitt98}). The explicit formula provided in Eq. \eqref{laplaceF} provides a simple and easy way to account for the distribution of dry periods with only 2 parameters, namely the integral scale $T$ and the dimension $D$.
In Fig. \ref{fig:rain4} this analytical formula provided by our model is compared 
to the empirical observations for two different months: March
which is a quite wet period where we estimated a fractal dimension $D \simeq 0.62$ and July which corresponds to a dry period with an estimated 
fractal dimension $D \simeq 0.47$. In both cases we assumed that the integral 
scale $T$ corresponds to 5 days. In both situations (that correspond to top 
and bottom panels of Fig. \ref{fig:rain4}) one can see that the analytical
curves ($\bullet$) perfectly fit the empirical data (solid lines). One notably 
sees in the insets that this very good agreement holds for the exponential 
tail  (insets with semi-logarithmic representation) as well as for the power-law 
behavior around the origin (insets with double-logarithmic representation).

\section{Summary and prospects}
\label{sec:conclusion}
In this paper we have introduced a new model for random cascades that live on 
random Cantor sets. 
This model can be considered as an extension of former constructions
of continuous cascades that accounts for a support with arbitrary Hausdorff (or capacity) dimension $d_C \in (0,1]$.
Basically these random measures are obtained as the product of a random cascade density $e^{\omega_\ell(t)}$ by a correlated 
Bernoulli random process $\delta_{\gamma_\ell(t)}$
that corresponds to a self-similar version of ``random cutouts" introduced by Mandelbrot more than forty years ago.
We have shown that the limit of our construction is well defined and possesses stochastic self-similarity properties with an arbitrary log-infinitely divisible multifractal spectrum $\zeta_q$.
We have also studied the distribution of void sizes that turns out to be a power-law truncated exponentially: at small scales, this distribution is a power  law $\tau^{-D}$ while at large scales it decreases exponentially with a characteristic size $\frac{T}{\alpha_D}$ 
related to both $T$, the integral scale of the model, and its dimension $D$.
Most of our analytical results have been illustrated (and checked) by various numerical simulations.
As a first application, we considered high resolution time series corresponding
to the precipitation intensity recorded during 5 years in 5 different locations in south west of France. Besides the basic multiscaling and fractal properties of the series, we have shown that our model provides a remarkable fit of the distribution of dry period durations. A specific application of this 
model to describe various statistics associated with rainfall variations will be considered in a future work.

Beyond applications to peculiar experimental 
situations, various questions remain open at
both mathematical and statistical levels. For example
one can consider
the extension of our construction 
to higher dimension, the questions related to 
subordination, the relationship between the Levy
process we introduced in section \ref{sec:void} 
and the so-called ``tempered
stable processes" \cite{RosTemp}. From a statistical 
point of view, a parametric estimation method 
of both $D$ and $T$ can be introduced and compared 
to the non-parametric method based on scaling as used in section \ref{sec:examples}.

\appendix

\section{Infinitely divisible measures in the $(t,s)$
half-plane}
\label{cone_app}
In this section we just recall the main lines of the 
construction proposed in Ref. \cite{MuBa02,BaMu03}
that involves some infinitely divisible measure 
$dm(t,s)$ spread over the half-plane $(t,s) \in {\mathbb R} \times {\mathbb R}^{+\star}$.
Let us consider a set ${\cal A} \subset {\mathbb R} \times {\mathbb R}^{+\star}$ and denote its ``area'', i.e.
 its measure as respect to the natural measure $s^{-2} dt ds$, as:
\begin{equation}
\label{area}
 S({\cal A}) = \int_{{\cal A}} s^{-2} dt ds \; .
\end{equation}
The random measure $dm(t,s)$ allows one to associate
with each set ${\cal A}$ an infinitely divisible 
random variable $m({\cal A}) = \int_{{\cal A}} dm(t,s)$
which characteristic function reads:
\begin{equation}
  \EE{e^{p m({\cal A})}} = e^{\psi(p) S({\cal A})}
\end{equation}
where $\psi(p)$ is the cumulant generating function of an infinitely divisible 
law as described by the Levy-Khintchine formula.

Of particular interest are the cone-like sets 
\begin{equation}
\label{def-cone}
  \cC_\ell(t_0) = \left\{(t,s), s \geq \ell, |t-t_0| < \frac{\min(T,s)}{2} \right\}
\end{equation}
where $0 < \ell < T$.

The following formula can be established by a straightforward computation 
and is useful in many results of this paper:
If $0 < \ell' \leq \ell$, one has,
\begin{equation}
\label{rhoexact}
\rho_{\ell}(\tau) \stackrel{def}{=} S \left[\cC_\ell(t) \cap \cC_{\ell'}(t+\tau)\right] \\ =
\left\{
\begin{array}{ll} 
\ln \left(\frac{T}{\ell} \right)+1-\frac{\tau}{\ell} & \mbox{if}~\tau \leq \ell \; , \\ 
\ln \left(\frac{T}{\tau} \right) & \mbox{if}~T \ge  \tau \ge \ell \; , \\
0 & \mbox{if}~ \tau > T \; .
\end{array}
\right. 
\end{equation} 
Notice that $\rho_\ell(0) = S\left[\cC_\ell(t)\right] = 1+\ln(\frac{T}{\ell})$ and that $\rho_\ell$ 
satisfies the remarkable property:
\begin{equation}
\label{ss-rho}
\rho_{s\ell}(st) = \rho_\ell(t)-\ln(s) \; .
\end{equation}
One can also define $\nu_{\ell,\ell'}(\tau)$ as:
\begin{equation}
\label{def-nu}
\nu_{\ell,\ell'}(\tau) \stackrel{def}{=} S \left[ \cC_\ell(t) \triangle \cC_{\ell'}(t+\tau)\right]
\end{equation}
where $\triangle$ stands for the symmetric difference between the two sets.
The expression of $\nu_{\ell,\ell'}(\tau) $ can be easily deduced from 
Eq. \eqref{rhoexact} since $\nu_{\ell,\ell'}(\tau) = 	\rho_{\ell}(0)+\rho_{\ell'}(0)-2 \rho_{\ell}(\tau)$.  

\section{The asymptotic behavior of the void size distribution}
\label{App:void_asympt}
In this section we study the asymptotic behavior of $F(\tau)$, the inverse Laplace transform 
of the expression \eqref{laplaceF} using Tauberian theorems.
Notice that Eq. \eqref{laplaceF} involves only the variable $sT$ so that $F(\tau)$ is 
a function of $\frac{\tau}{T}$. With no loss of generality one can then set $T=1$.

\begin{center}
	\begin{figure}[h]
		\includegraphics[width=0.6 \textwidth]{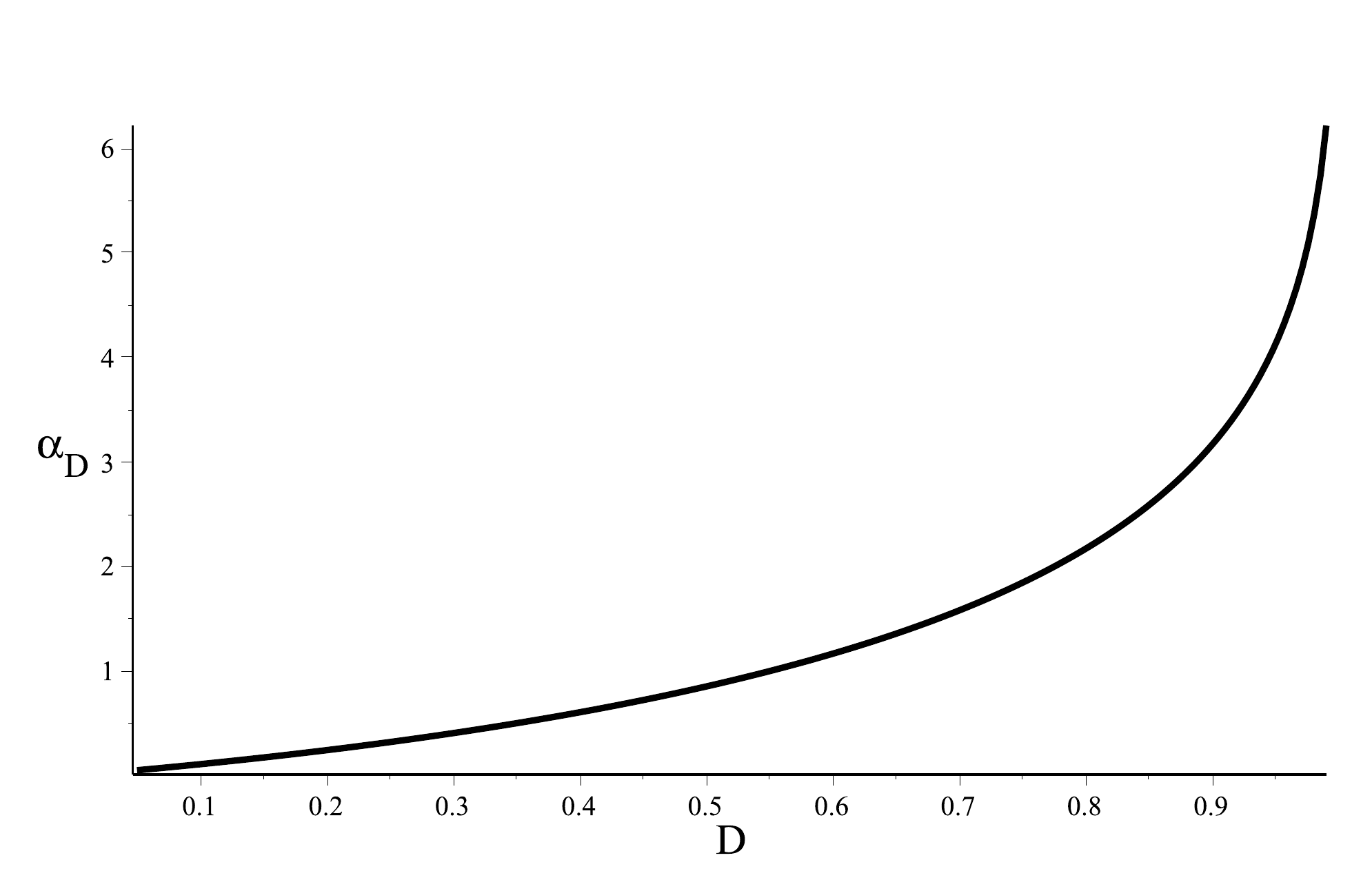}
		\caption{The asymptotic exponential decreasing rate $\alpha_D$ (the opposite of the solution of Eq. \eqref{pole_eq}) as a function of the dimension $D$.}
			\label{fig:alpha_D}
	\end{figure}
\end{center}

In the limit $z \to 0$, one can directly use Theorem 3 and Theorem 4 in Sec. XIII.5 of \cite{Fel71} that link 
the behavior of the Laplace transform when $\alpha \to \infty$ to the behavior of the function around 
the origin.
Since $\gamma(x,t) \to \Gamma(x)$ when $t \to \infty$, we have from \eqref{laplaceF}:
\begin{equation}
  \int_0^{\infty} \! \! \! e^{-\alpha z} F(z) dz \operatornamewithlimits{\sim}_{\alpha \to \infty}  \frac{\alpha^{D-1}}{\Gamma(D)} \; .
 \end{equation}
It then results that
\begin{equation} 
  F(\tau) \operatornamewithlimits{\sim}_{\tau \to 0} \tau^{-D}  \; .
\end{equation}
Since $\frac{\phi(\alpha)}{\alpha} \to 1$ when $\alpha \to 0$, one can no longer invoke classical 
Tauberian theorems of Ref. \cite{Fel71} but one can use the results of Ref. \cite{TailLaplace} in 
which the author proves that the largest real part of the poles of the Laplace transform directly provides 
the exponential rate of the tail behavior of a function. It follows that
$$
 F(\tau) \operatornamewithlimits{\sim}_{\tau \to \infty} e^{-\alpha_D \tau}
$$
where $-\alpha_D$ is the largest (negative) real part of the poles of the Laplace transform of $F(\tau)$.
In our case this value corresponds to solutions of the equation:
\begin{equation}
    z^{1-D} \gamma(D,z)+e^{-z} = 0
\end{equation}
that can be rewritten in terms of confluent hypergeometric function as 
\begin{equation}
\label{pole_eq}
_1F_1(1,D+1,z) + \frac{D}{z} = 0 \: .
\end{equation}
This transcendental equation cannot be solved analytically but numerical solutions are easy
to obtain. Moreover one can prove that $\alpha_D \sim D$ when $D \to 0$ while 
$\alpha_D \to \infty$ when $D \to 1$.
The estimated values of $\alpha_D$ as a function of $D$ are reported in Fig. \ref{fig:alpha_D}.

%
%
%

\section{Proof of the convergence of $\mu_\ell$}
\label{conv_app}
In this section, we prove that, if $\psi(1) = 1-D$ and $\psi(2)< 2-D$, then 
$\mu_\ell(t) = \int_0^t d\mu_\ell(u)$ where
\begin{equation}
d\mu_\ell(u) = du \; e^{\omega_\ell(u)} \delta_{\gamma_\ell(u)} 
\end{equation}
converges in the mean square sense when $\ell \to 0$.
We follow the same line as in ref. \cite{MuBa02} and 
define
\begin{equation}
R_{\ell,\ell'}(\tau) = \EE{ e^{\omega_\ell(u)+\omega_{\ell'}(u+\tau)}\delta_{\gamma_\ell(u)}\delta_{\gamma_{\ell'}(u+\tau)}   } \; .
 \end{equation}
Let us first show that, if $\ell' \leq \ell$:
\begin{equation}
\label{rl}
R_{\ell,\ell'}(\tau)  = R_{\ell,\ell}(\tau) = e^{\rho_\ell(\tau)(\psi(2)+D-1)} 
\end{equation}
with $\rho_{\ell}(\tau)$ as defined in Eq. \eqref{rhoexact}).

Because $\gamma_\ell$ and $\omega_\ell$ are independent, we have
\begin{equation}
\label{rl2}
R_{\ell,\ell'}(\tau) = \EE{ e^{\omega_\ell(u)+\omega_{\ell'}(u+\tau)}} 
\EE{\delta_{\gamma_\ell(u)}\delta_{\gamma_{\ell'}(u+\tau)}} \; .
\end{equation}

From the definition \eqref{def-omega} of $\omega_\ell$, one has
\begin{equation}
\omega_\ell(u)+\omega_{\ell'}(u+\tau) =  \int_{\cC_\ell(u) \cup \cC_\ell'(u+\tau)} \! \! \! \! \! \! \! \! dm(t,s)  
 =  \int_{\cC_\ell(u) \triangle \cC_\ell'(u+\tau)}  \! \! \! \! \! \! \! \! dm(t,s) +2 \int_{\cC_\ell(u) \cap \cC_\ell'(u+\tau)}  \! \! \! \! \! \! \! \! \! dm(t,s)
\end{equation}
where $\triangle$ stands for the symmetric difference of two sets.
The last two terms (denoted as $\omega_s$ and $\omega_i$) involve
the integral of $dm(t,s)$ over disjoint intervals and are therefore
independent random variables.
From the definition of the functions $\rho_\ell(\tau)$ and 
$\nu_{\ell,\ell'}(\tau)$ of Appendix \ref{cone_app}, one then has:
\begin{eqnarray}
\EE{ e^{\omega_\ell(u)+\omega_{\ell'}(u+\tau)}} & = & \EE{e^{\omega_s}}  \EE{e^{\omega_i}} \\ 
\label{f1}
& = & e^{\psi(1) \nu_{\ell,\ell'}(\tau) }e^{\psi(2) \rho_\ell(\tau)} \; . 
\end{eqnarray}

Estimating the second factor in Eq. \eqref{rl2} just amounts to computing the probability that
both $\gamma_\ell(u)$ and $\gamma_\ell'(u+\tau)$ vanish, i.e. the probability that
there is no Poisson event in the domain $\cC_\ell(u) \cup \cC_\ell(u+\tau) = \left(\cC_\ell(u) \triangle \cC_\ell(u+\tau)\right) \cup \left(\cC_\ell(u) \cap \cC_\ell(u+\tau)\right)$.  Since the Poisson density is $dp(t,s) = (1-D) s^{-2} dt ds$,
the probability that there is no event in some set ${\cal E}$ in the $(t,s)$ plane is simply 
$e^{-\int_{\cal E} dp(t,s)}= e^{(D-1)S({\cal E})}$.
Along the same line as above, using Eqs. \eqref{rhoexact} and \eqref{def-nu} of Appendix \ref{cone_app}, we thus have:  
\begin{equation}
\label{f2}
\EE{\delta_{\gamma_\ell(u)}\delta_{\gamma_{\ell'}(u+\tau)}} = e^{(D-1)[\rho_\ell(\tau)+\nu_{\ell,\ell'}(\tau)]}  \; .
\end{equation}
Since, from Eq. \eqref{condpsi}, we have assumed that $\psi(1) = 1-D$,
the product of \eqref{f1} and \eqref{f2} leads to Eq. \eqref{rl}. 

Our goal is to show, that,
$\forall \; \epsilon$, $\exists \; \ell_0$, $\forall \; \ell,\ell' < \ell_0$,
$\EE{ (\mu_\ell(t)-\mu_{\ell'}(t))^2 } < \epsilon$.
Let us suppose that $\ell'\leq \ell$.
Then,
\begin{eqnarray*}
	& & \EE{ (\mu_\ell(t)-\mu_{\ell'}(t))^2} =  \\
	& & \EE {\mu_\ell^2(t)}+\EE{\mu_{\ell'}^2(t)}
	-2 \EE{ \mu_\ell(t) \mu_{\ell'}(t)} = \\\
	& &  \int_0^t \! \! \int_{0}^{t} \! \! \left( R_{\ell,\ell}(u-v) \! +  \! R_{\ell',\ell'}(u-v) \!- \!2 R_{\ell,\ell'}(u-v) \right)  du dv  \; .
\end{eqnarray*}
Thus, thanks to Eq. (\ref{rl}) and the expression \eqref{rhoexact} of $\rho_\ell$, we get, after a little algebra
\begin{eqnarray*}
	\EE{ (\mu_\ell(t)-\mu_{\ell'}(t))^2} & = &  
	t \left( \int_{0}^{t} \left[ R_{\ell',\ell'}(u)-R_{\ell,\ell}(u) \right] du \right) \\
	& = & O \left( \ell^{2-D-\psi(2)} \right)  \; .
\end{eqnarray*}
Thus if $\psi(2) < 2-D$, we see that $\mu_\ell(t)$ is a 
Cauchy sequence and thus converges in mean square sense. 
The same type of argument also proves the mean-square convergence (and therefore the convergence in law) 
of all finite dimensional vectors $(\mu_\ell([t_1,t_1+\tau_1]),\ldots,\mu_\ell([t_n,t_n+\tau_n]))$.
In order to achieve the proof of the weak convergence we must establish a tightness
condition. For that purpose, it suffices to bound some 
moment of $\mu(t)$. Such a bound is directly obtained
for the second moment using Eq. \eqref{rl}:
\begin{equation}
\EE{\mu_\ell(t)^2} = t \int_0^t R_{\ell,\ell}(\tau) \leq C t^{3-D-\psi(2)} 
\end{equation}
where $C$ is a constant that does not depend on $\ell$.

\section{Proof of the self-similarity of $\mu_\ell$}
\label{ss_app}
In order to prove the stochastic self-similarity of $\mu(t)$, according to section \ref{sec:sss}, one just needs to 
prove Eqs. \eqref{ss-1} and \eqref{ss-2}.
Since \eqref{ss-1} is already proven in \cite{BaMu03}, it just remains to show that:
\begin{equation}
\label{ss-gamma}
 \gamma_{s \ell}(s u)  \stackrel{law}{=}  \Gamma_s + \gamma_\ell(u)
\end{equation}
where $\Gamma_s$ is a random variable independent of the process $\gamma_\ell(t)$ with the same Poisson law as 
$\gamma_{s T}(t)$.
This result can be directly deduced from Lemma 1 of \cite{BaMu03}. Indeed this lemma gives 
the  $q$-point characteristic function of the infinitely divisible process $\gamma_\ell(t) = \int_{\cC_\ell(t)} dp(v,s)$ 
as defined  by Eq. \eqref{defgamma}.
Let $\varphi(p)= \psi(-ip) = (1-D) (e^{ip}-1)$ be the cumulant generating function associated with the Poisson process $dp(v,s)$
(see the definition in Eq. \eqref{defpsi} of the cumulant generating function associated with some infinitely divisible measure in the 
$(t,s)$ plane); let $q \in \bm{N}^*$, $\vec{t}_{q} = (t_1,t_2,\ldots,t_q)$ with $t_1 \le t_2 \le \ldots \le t_n$ and $\vec{p}_{q} = (p_1,p_2,\ldots,p_q)$. 
Then, the characteristic function of the vector $\{\gamma_\ell(t_m)\}_{1\le m\le q}$ reads:
\begin{equation}
\label{char}
\EE{e^{\sum_{m=1}^q i p_m \gamma_{\ell}(t_m)}}  = e^{\sum_{j=1}^q \sum_{k=1}^{j}
	\alpha(j,k) \rho_\ell(t_k-t_j)}
\end{equation}
where $\rho_\ell(t)$ is given by Eq. \eqref{rhoexact} and the coefficients $\alpha(j,k)$ provided
in \cite{BaMu03} satisfy:
\begin{equation}
\label{rk}
\sum_{j=1}^q \sum_{k=1}^{j}\alpha(j,k) = \varphi \left( \sum_{k=1}^q p_k \right)  \; .
\end{equation}
From Eq. \eqref{char}, \eqref{ss-rho} and \eqref{rk}, we have therefore,
\begin{equation}
\EE{e^{\sum_{m=1}^q i p_m \gamma_{s \ell}(s t_m)}}  = \EE{e^{\sum_{m=1}^q i p_m \gamma_{s \ell}(s t_m)}} e^{-\ln(s) \varphi \left(\sum_{k=1}^q p_k \right)} 
\end{equation}
which is equivalent to Eq. \eqref{ss-gamma}.

\end{document}